\documentclass[aps,prb,twocolumn,superscriptaddress]{revtex4-1}
\usepackage{natbib}
\usepackage{graphicx,amssymb,amsmath,ifthen}
\usepackage{bm}

\begin{document}


\title{Two terminal charge tunneling: Disentangling Majorana zero modes from partially separated Andreev bound states in semiconductor-superconductor heterostructures}


\author{Christopher Moore}
\affiliation{Department of Physics and Astronomy, Clemson University, Clemson, SC 29634, USA}

\author{Tudor D. Stanescu}
\affiliation{Department of Physics and Astronomy, West Virginia University, Morgantown, WV 26506, USA}

\author{Sumanta Tewari}
\affiliation{Department of Physics and Astronomy, Clemson University, Clemson, SC 29634, USA}



\date{\today}

\begin{abstract}
We show that a pair of overlapping Majorana bound states (MBSs) forming a partially-separated Andreev bound state (ps-ABS) represents a generic low-energy feature in spin-orbit coupled semiconductor-superconductor (SM-SC) hybrid nanowire in the presence of a Zeeman field. In a finite nanowire the ps-ABS interpolates continuously between the ``garden variety'' ABS, which consists of two MBSs sitting on top of each other, and the topologically protected Majorana zero modes (MZMs), which are separated by a distance given by the length of the wire. The really problematic ps-ABSs consist of component MBSs separated by a distance of the order of the characteristic Majorana decay length $\xi$, and have nearly zero energy in a significant range of control parameters, such as the Zeeman field and chemical potential, within the topologically trivial phase. Despite being topologically trivial, such ps-ABSs can generate signatures identical to MZMs in local charge tunneling experiments. In particular, the height of the zero bias conductance peak (ZBCP) generated by ps-ABSs has the quantized value, $2e^2/h$, and it can remain unchanged in an extended range of experimental parameters, such as Zeeman field and the tunnel barrier height.  We illustrate the formation of such low-energy robust ps-ABSs in two experimentally relevant situations: a hybrid SM-SC system consisting of a proximitized nanowire coupled to a quantum dot and the SM-SC system in the presence of a spatially varying inhomogeneous potential. We then show that, unlike local measurements, a two-terminal experiment involving charge tunneling at both ends of the wire is capable of distinguishing between the generic ps-ABSs and the non-Abelian MZMs. While the MZMs localized at the opposite ends of the wire generate correlated differential conduction spectra, including correlations in energy splittings and critical Zeeman fields associated with the emergence of the ZBCPs, such correlations are absent if the ZBCPs are due to ps-ABSs emerging in the topologically trivial phase. Measuring such correlations is the clearest and most straightforward test of topological MZMs in SM-SC heterostructures that can be done in a currently accessible experimental set-up.
\end{abstract}

\pacs{}

\maketitle

\section{Introduction}

Topological superconductors are characterized by a gap in the bulk excitation spectrum and gapless excitations, called Majorana modes, on the boundary.\cite{Read_Green_2000,Kitaev_2001,Nayak_2008,Beenakker,Franz} The zero energy Majorana modes, or Majorana zero modes (MZMs), localized near topological defects or the edges of one-dimensional systems are predicted to have non-Abelian exchange statistics and have been proposed as building blocks for topological qubits.
While MZMs have not yet been conclusively observed in experiments, they have been theoretically shown to exist in low-dimensional spinless $p$-wave superconductors,\cite{Read_Green_2000,Kitaev_2001} in topological insulators with proximity induced superconductivity,\cite{Fu_2008} and more recently in low-dimensional spin-orbit-coupled semiconductor-superconductor (SM-SC) heterostructures in the presence of a Zeeman field.\cite{Sau,Annals,Alicea,Long-PRB,Roman,Oreg,Stanescu}
The SM-SC heterostructure,\cite{Sau,Annals,Alicea,Long-PRB,Roman,Oreg,Stanescu} which  involves a low-dimensional semiconductor with spin-orbit coupling in proximity to an s-wave superconductor in the presence of an applied Zeeman field,
has motivated tremendous experimental efforts in the last few years.\cite{Mourik_2012,Deng_2012,Das_2012,Rokhinson_2012,Churchill_2013,Finck_2013,Marcus,Hansen,HZhang,Frolov,Nichele,Quantized-ZBCP} It was shown theoretically that the applied Zeeman field $\Gamma$ drives the SM-SC heterostructure through a
topological quantum phase transition (TQPT) at a critical Zeeman field $\Gamma=\Gamma_c$ to a topologically
non-trivial superconducting phase with a zero-energy MZM localized at each end of the wire.
While  for $\Gamma < \Gamma_c$ the system is topologically trivial and characterized (in principle) by a finite quasiparticle gap,
for $\Gamma > \Gamma_c$, i.e.  in the topologically non-trivial phase,   the edge-localized MZMs  are expected to be
observable in local charge tunneling experiments at the ends of the wire as a quantized zero bias conductance
peak (ZBCP) with a peak height (at zero temperature) of $2e^2/h$.\cite{Sengupta,Akhmerov,Law,Flensberg}
In the past few years several observations of robust ZBCPs consistent with the presence of MZMs in quasi-one-dimensional SM-SC heterosructures in the presence of a Zeeman field have been reported in the literature \cite{Mourik_2012,Deng_2012,Das_2012,Churchill_2013,Finck_2013,Marcus,Hansen,HZhang,Frolov,Nichele,Quantized-ZBCP}. Recently, the fabrication of high-quality semiconductor wire - epitaxial superconductor structures has allowed the measurement of charging effects in the Coulomb blockade regime \cite{Marcus}. Measurements of the zero-bias conductance in charge tunneling with normal leads attached at each end of the wire (the so-called teleportation signal \cite{Fu}), suggest a suppression of the lowest mode energy splitting with increasing wire length, which was cited as evidence for the exponential topological protection of MZMs \cite{Marcus,Heck}. More recently, a quantized conductance peak of height $2e^2/h$ was found in InSb nanowires covered with superconducting Al. The quantization of the peak height was found to be robust to variations in the external parameters, such as the Zeeman field and the height of the tunnel barrier. \cite{Quantized-ZBCP}

	
It was shown earlier that trivial low-energy sub-gap states can emerge rather generically in semiconductor-superconductor hybrid systems in the presence of disorder\cite{Bagrets,Liu,Sen1,Sen2,Loss,Wimmer}, non-uniform system parameters\cite{Kells,Bena,Roy,Jose,Ojanen,Stanescu-Tewari,Aguado,Klinovaja,Pablo-San-Jose,fleckenstein}, weak antilocalizaton\cite{Pikulin}, or coupling to a quantum dot\cite{Prada,Liu-Sau}. In many situations these low-energy states are found to generate ZBCPs that are not quantized at $2e^2/h$ and/or are not robust against variations of control parameters such as magnetic field, chemical potential, and  tunnel
barrier height. One may be tempted to assume that this property of the trivial low-energy states is generic and that, consequently, one can distinguish them from topologically-protected MZMs, which generate similar signatures in a tunneling conductance measurement but produce quantized ZBCPs that are robust to variations in the local parameters. For instance, in a recent theoretical work,\cite{Liu-Sau} where the experiment of Ref.~[\onlinecite{Hansen}] was modelled as a proximitized nanowire coupled to a metallic lead via a quantum dot (junction composed of lead and quantum dot-nanowire-superconductor heterostructure with the quantum dot subject to a confining potential \cite{Hansen,Prada,Liu-Sau}), it was claimed that the ZBCP due to ABS in the topologically trivial phase is not stable and oscillates (or splits) as a function of the dot potential, and this property can be used to distinguish between low energy ABSs in the topologically trivial phase and MZMs in the topologically non-trivial phase, which are robust to such local variation in the dot potential. However, as pointed out recently\cite{Stanescu-Tewari-2} and demonstrated in detail in this paper, this is not generically the case, since there are situations involving partially separated ABSs (ps-ABS) where one cannot distinguish between trivial low-energy modes and MZMs using \textit{any} type of local measurement at the end of the wire. This includes the property that the height of the ZBCPs generated by both MZMs and ps-ABSs are quantized (to the value $2e^2/h$) over an extended range of parameters.
We note that, while the emergence of trivial low-energy sub-gap states has been investigated in the literature, \cite{Bagrets,Liu,Sen1,Sen2,Loss,Wimmer,Kells,Bena,Roy,Jose,Ojanen,Stanescu-Tewari,Aguado,Klinovaja,Pablo-San-Jose,fleckenstein,Prada,Liu-Sau}
the introduction of the notion of ps-ABSs as {\em generic robust low-energy features} in SM-SC heterostructures with non-uniform system parameters [e.g., quantum dot-nanowire-superconductor heterostructures (Sec.~III) and systems with long length scale potential inhomogeneity (Sec.~IV)] and the demonstration that such modes can produce quantized ZBCPs in local tunneling experiments over an extended range of parameters, virtually identical to the corresponding signatures of topological MZMs, are new contributions of this work that were not anticipated before.

To gain further insight into the possible types of behavior exhibited by the trivial low-energy ABSs, we represent such a state as a pair of  Majorana bound states (MBSs), each of which can be represented by a second quantized operator satisfying the Majorana condition $\gamma^{\dagger}=\gamma$. Such a construction (Eq.~\ref{chiA} and Eq.~\ref{chiB} below) can obviously be done for any eigenstate of the Bogoliubov-de Gennes (BdG) Hamiltonian, but in this paper we are interested only in the low energy states that appear in the sub-gap energy range in SM-SC heterostructures at finite values of the Zeeman field. As we show below, the ps-ABS is a convenient concept that interpolates continuously between the ``garden variety'' ABS, which consists of two MBSs sitting ``on top of'' each other, and the topologically protected pair of MZMs, which are separated by a distance given by the length of the wire ($L$). Such an interpretation of an ABS in
the crossover regime between small and large overlap of
component MBSs was used earlier \cite{Haim_prl}, and argued to produce non-universal behavior in tunneling experiments \cite{Haim}, but their importance for interpreting the existing ZBCP data in SM-SC heterostructures in the presence of Zeeman field was not discussed.

 The really problematic ps-ABSs consists of MBSs separated by a distance of the order of the characteristic Majorana length-scale ($\xi$) or larger (but less than the length of the wire), in which case the near zero energy of such a state can be robust to changes in a multitude of experimental parameters, e.g., Zeeman field, chemical potential, barrier height, induced pair potential, and, when a quantum dot is present,\cite{Hansen,Prada,Liu-Sau} the dot potential. When one of the constituent MBSs of such a ps-ABS  is located at the end of the wire, it generates experimental signatures in charge tunneling measurements (e.g., local charge tunneling into one end of the wire, resonant charge tunneling between the two ends in the Coulomb blockade regime \cite{Marcus}) that are identical to those produced by a ``true'' MZM which is separated from the other MZM by the length of the wire. Furthermore, the height of the ZBCP generated by such a ps-ABS has the quantized value, $2e^2/h$, while the peak itself is robust to an extended range of Zeeman field, chemical potential, and other experimental parameters despite being topologically trivial (i.e., appearing in the topologically trivial phase for $\Gamma < \Gamma_c$).

We will explicitly show below that ps-ABSs -- irrespective of whether they are produced in a lead-quantum dot-nanowire-superconductor set up \cite{Hansen,Prada,Liu-Sau} or proximitized nanowire with long-length-scale potential inhomogeneity -- cannot be distinguished from topological MZMs by any local measurements (e.g., one-terminal charge or spin tunneling measurements). On the other hand, experiments involving quasiparticle interference, fusion, or braiding (which should be able to discriminate between trivial ps-ABSs and topological MZMs)  are expected to be technically difficult.\cite{Sau_Swingle_Tewari,Dahan,Sau-Interference,Hell}  Interpreting the results of such an experiment without knowing whether or not the system harbors ps-ABSs or just a single pair of  MZMs localized at the opposite ends of the wire would be a complicated task. Therefore, before conducting such an experiment, it may be crucial to perform simpler tests that are sensitive to the presence of ps-ABSs, and are able to distinguish between robust ZBCP signatures from the topologically trivial and non-trivial regimes. We will show below that the most straightforward such test, one which may be well within the capabilities of current experiments, is a two-terminal charge tunneling measurement.
Such an experiment involving tunneling into the two ends of the wire was proposed before,\cite{Sankar} but its ability to discriminate between topological MZMs and trivial low-energy states was not discussed explicitly. Such a discussion is particularly important in the new context of
 ps-ABSs, which are generic robust low energy features in non-uniform SM-SC heterostructures that cannot be distinguished from topological MZMs by any {\em local} measurement. Here, we show explicitly (using model calculations) that such as experiment can, in fact, discriminate between ps-ABSs and topological MZMs. Since the absence of non-uniformities (e.g., unintentional quantum dots) in proximitized wires has not been demonstrated experimentally (which leaves open the possibility of actually having observed ps-ABSs, rather than MZMs), our results reinforce the urgency of performing the two-terminal measurement, the most straightforward non-local measurement involving a type of experimental set-up that is currently accessible.

The two-terminal charge tunneling experiment involves collecting two sets of data: the first set records the differential conductance at the left end of the wire by applying a bias potential to the left lead, the superconductor being grounded, and the right lead being isolated; the second set, on the other hand, collects the differential conductance at the right end by applying a bias potential to the right lead, the superconductor being grounded and the left lead being isolated. The tunneling potentials as well as all other gate potentials are kept fixed throughout the measurement of both sets of experimental data. We will show in this paper that comparing the left and the right lead differential conductances, $(dI/dV)_L$ and $(dI/dV)_R$, provides essential information about the origin of the ZBCPs, being able to distinguish between ps-ABSs and topological MZMs from the topologically trivial and non-trivial phases, respectively. While the MZMs in the topological superconducting phase ($\Gamma > \Gamma_c$) will be characterized by correlated differential conduction spectra in both terminals, including correlations in energy splittings and critical Zeeman fields associated with the emergence of ZBCPs, such correlations will be absent if the ZBCPs are due to ps-ABSs in the topologically trivial phase ($\Gamma < \Gamma_c$).

The rest of the paper is organized as follows: In Sec.~\ref{Partial} we define the concept of ps-ABS using BdG Hamiltonian and low energy eigenstates. In Sec.~\ref{quantum-dot} and Sec.~\ref{Inhomogeneity} we illustrate the emergence of low energy robust ps-ABSs in the topologically trivial regime in lead-quantum dot-nanowire-superconductor heterostructures and promitized nanowire with long length scale potential variations, respectively. In Sec.~\ref{Probe} we show that measurements of ZBCP from one end of the wire or even two-terminal sub-gap resonant charge tunneling in the presence of Coulomb blockade are unable to distinguish between ps-ABS and MZMs, as they produce nearly identical signatures.
In Sec.~\ref{Two-terminal} we show that, in the absence of an interferometric experiment, a two-terminal set-up measuring the presence or absence of correlations in the tunneling spectra from the two ends should be able to reliably discriminate between ps-ABSs and topological MZMs. We end with a summary and several conclusions in Sec.~\ref{Conclusion}.

\section{Partially separated Andreev bound states}
\label{Partial}
To define the ps-ABS more precisely, let us start
with the Bogoliubov-de Gennes (BdG) Hamiltonian of the proximitized semiconductor coupled to a superconductor as given by,
\begin{eqnarray}\label{eq:7}
	&H_0 =-t_x\sum_{i,\delta_x,\sigma}c_{i+\delta_x\sigma}^\dagger
	c_{i\sigma}-t_y\sum_{i,\delta_y,\sigma}c_{i+\delta_y\sigma}^\dagger
	c_{i\sigma}\nonumber\\
	&-\mu\sum_{i,\sigma}c_{i\sigma}^{\dagger}c_{i\sigma}+ \Gamma\sum_{i,\sigma\sigma^\prime}c_{i,\sigma}^{\dagger}\tau_{\sigma\sigma^\prime}^x c_{i,\sigma^\prime} \nonumber \\
	& +\frac{i}{2}\sum_{i,\delta,\sigma\sigma^\prime}\left[\alpha c_{i+\delta_x,\sigma}^{\dagger}\tau_{\sigma\sigma^\prime}^y c_{i,\sigma^\prime}-\alpha_y c_{i+\delta_y,\sigma}^{\dagger}\tau_{\sigma\sigma^\prime}^x c_{i,\sigma^\prime} +h.c.\right]\nonumber\\ &+\sum_{i}\Delta_{ind}\left(c_{i\uparrow}^{\dagger}c_{i\downarrow}^{\dagger}+h.c.\right)\tau_x,
\end{eqnarray}
where the lattice sites $i$ correspond to $N_y$ parallel chains oriented along the $x$-direction, $\tau^x$ and $\tau^y$ are $2\times 2$ Pauli matrices acting in the spin space, $\sigma$ and $\sigma^\prime$ are spin indices, $t_x$ and $t_y$ are hopping matrix elements, $\delta_x$ and $\delta_y$ are nearest neighbor vectors, $\mu$ is the chemical potential, $\Gamma$ the Zeeman field, $\alpha$ and $\alpha_y$ the longitudinal and transverse Rashba coeficients, respectively, and $\Delta_{ind}$ is the pair potential in the semiconductor proximity induced from a host superconductor. In this paper we take $N_y =1$, the case of a single chain, leading to $t_y =\alpha_y=0$. In Sec. \ref{quantum-dot}, while discussing the quantum dot-nanowire-SC heterostructure, we take $\Delta_{ind}$ to be non zero and uniform only in the part of the SM wire covered by the SC and zero outside, see Fig. I. In Sec. IV, while discussing the effect of long length scale variations, we supplement the Hamiltonian in Eq. I by a position dependent potential function $V_c (i)$ (see Eq. 6).
 Consider a low-energy solution of the effective BdG Hamiltonian $H_0$ corresponding to a positive energy $\epsilon \ll \Delta_{ind}$. Using the spinor representation $\psi_i = (c_{i\uparrow}, c_{i\downarrow}, c_{i\uparrow}^\dagger, c_{i\downarrow}^\dagger)^T$, where $i$ labels the position along the wire, one can write the wave function of the low-energy state in the form $\phi_\epsilon(i) = (u_{i\uparrow},   u_{i\downarrow}, v_{i\uparrow},   v_{i\downarrow})^T$. Particle-hole symmetry ensures that there will also be a negative-energy solution of the BdG equation described by the wave function $\phi_{-\epsilon}(i) = (v_{i\uparrow}^*,   v_{i\downarrow}^*, u_{i\uparrow}^*,   u_{i\downarrow}^*)^T$. Using these solutions, we construct the linear combinations
\begin{eqnarray}
\chi_A(i) &=& \frac{1}{\sqrt{2}}\left[\phi_{\epsilon}(i) +\phi_{-\epsilon}(i)\right], \label{chiA}\\
\chi_B(i) &=& \frac{i}{\sqrt{2}}\left[\phi_{\epsilon}(i) -\phi_{-\epsilon}(i)\right].     \label{chiB}
\end{eqnarray}
These states have a spinor structure of the form $\chi_\alpha(i)=(\widetilde{u}_{\alpha i\uparrow}, \widetilde{u}_{\alpha i\downarrow}, \widetilde{u}_{\alpha i\uparrow}^*, \widetilde{u}_{\alpha i\downarrow}^*)^T$, where $\alpha = A, B$ and $u_{A,i,\sigma}=u_{i\sigma} +v_{i\sigma}^*$, while $u_{B,i,\sigma}=i(u_{i\sigma} -v_{i\sigma}^*)$.
The corresponding creation operators,
\begin{eqnarray}
\gamma_\alpha^\dagger = \sum_i \left(\widetilde{u}_{\alpha i\uparrow}c_{i\uparrow}^\dagger + \widetilde{u}_{\alpha i\downarrow}c_{i\downarrow}^\dagger + \widetilde{u}_{\alpha i\uparrow}^*c_{i\uparrow} +  \widetilde{u}_{\alpha i\downarrow}^*c_{i\downarrow}\right),
\end{eqnarray}
manifestly satisfies the Majorana condition, $\gamma_\alpha^\dagger = \gamma_\alpha$. Hence, given the low-energy BdG states $\phi_{\epsilon}$ and $\phi_{-\epsilon}$, one can uniquely define the modes $\chi_A$ and $\chi_B$ corresponding to the Majorana operators $\gamma_A$ and $\gamma_B$. Note, however, that these Majorana modes are not exact eigenstates of the BdG Hamiltonian. More specifically, we have
\begin{eqnarray}
\langle\chi_\alpha|H|\chi_\alpha\rangle &=& 0, \nonumber \\
\langle \chi_A|H|\chi_B\rangle &=& i\epsilon.        \label{AHB}
\end{eqnarray}
Also, the construction described above can be done for any eigenvalue of the BdG Hamiltonian.
Nonetheless, it has special significance for the low-energy sub-gap states, which are typically localized.

\begin{figure}[t]
\begin{center}
\includegraphics[width=0.48\textwidth]{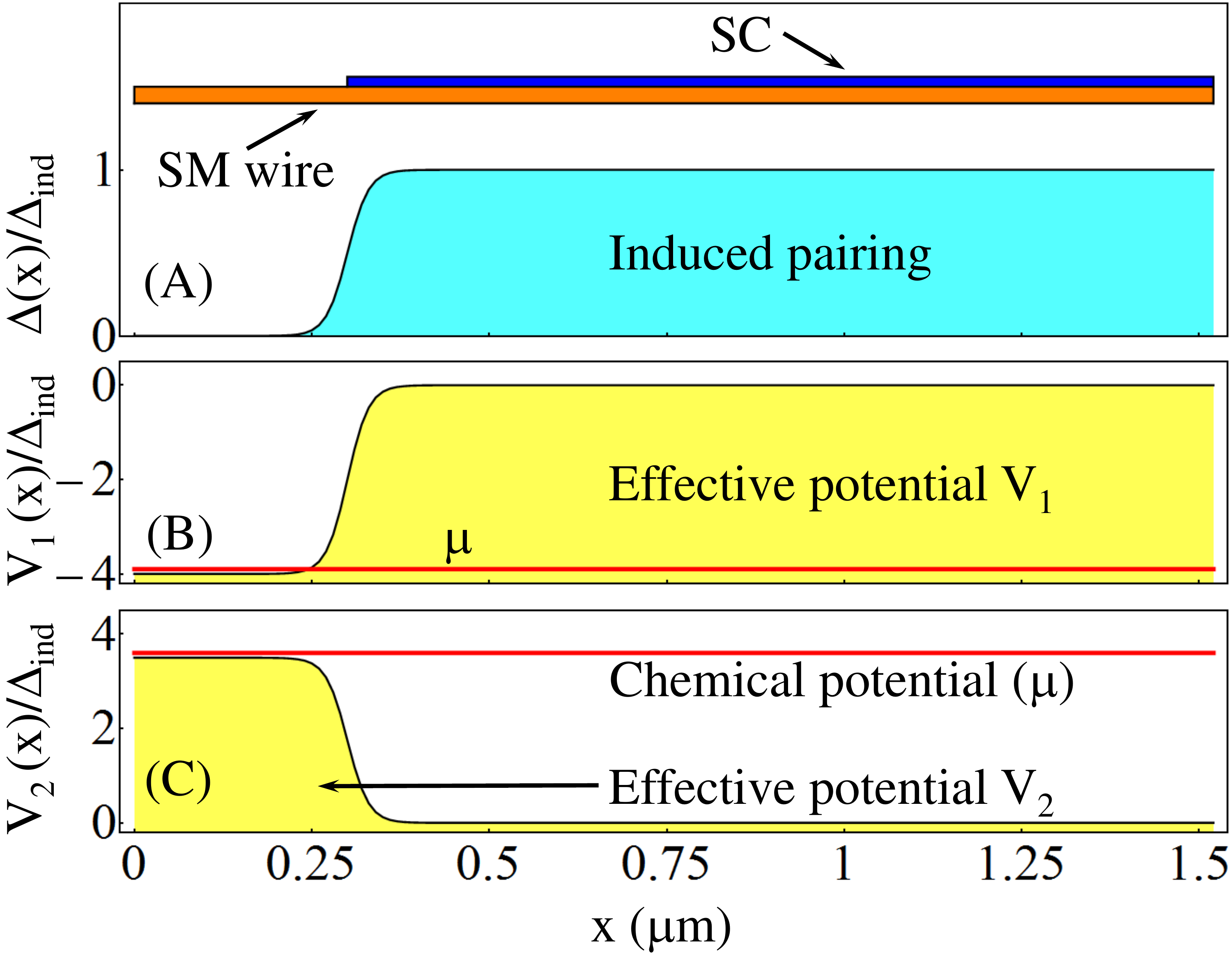}
\end{center}
\vspace{-2mm}
\caption{(Color online) Schematic representation of the SM-SC heterostructure in Ref.~[\onlinecite{Hansen}] (see also Ref.~[\onlinecite{Prada}] and Ref.~[\onlinecite{Liu-Sau}]). Proximitized semiconductor nanowire coupled to a quantum dot, where the dot is represented by the segment of the wire that is not covered by the superconductor. (A) The induced paring is $\Delta_{ind}=0.25~$meV in the proximitized region and vanishes in the dot region. (B) Effective potential $V_1(x)$ normalized by $\Delta_{ind}=0.25~$meV; the dot corresponds to a potential valley. (C)   Effective potential $V_1(x)$ normalized by $\Delta_{ind}=0.25~$meV; the dot is modeled as a potential step. In this paper we model the quantum dot as a potential step, while Refs.~[\onlinecite{Prada},\onlinecite{Liu-Sau}] model the quantum dot as a potential valley. Which one of the profiles $V_1(x)$ or $V_2(x)$ better represents the potential characterizing the quantum dot in the quantum dot-nanowire-superconductor  hybrid devices realized in the laboratory depends on the specific materials, the work function difference between the semiconductor and the superconductor,
and on the voltage applied under the dot region.}
\label{XFig1}
\vspace{-3mm}
\end{figure}

The spatial profiles of the Majorana bound states constructed above provide extremely useful information concerning the robustness of the low-energy mode $\phi_\epsilon$ with respect to variations of the control parameters. More specifically, strong overlap (i.e., spatial separations between $\chi_A$ and $\chi_B$ smaller than the characteristic length scale $\xi$) implies that $\epsilon$ is extremely sensitive to variations of the control parameters (e.g., Zeeman field) and, consequently, the low-energy mode $\phi_\epsilon$ will not ``stick'' to zero energy.  By contrast, well separated MBSs (i.e., a separation between $\chi_A$ and $\chi_B$ of the order of $\xi$ or larger) will result in a robust low-energy mode since the overlap in Eq. \ref{AHB} is small ``by construction'', regardless of the specific parameter values. In addition, these spatial profiles can provide information about the signature of the low-energy mode $\phi_\epsilon$ in a local measurement. If the couplings between the probe and the modes $\chi_A$ and $\chi_B$ are comparable, the signature will correspond to a standard ABS. If, on the other hand, the MBSs are well separated (the separation between them being $\sim \xi$ or larger) and the probe couples strongly to one of them and (exponentially) weakly to the other, the corresponding signatures will be indistinguishable from that of a ``true'' MZM.

Based on these general considerations, we distinguish three classes of low-energy modes that can emerge in semiconductor-superconductor hybrid systems.  First, we have the ``standard'' ABSs, which are low-energy modes that correspond to strongly overlapping MBSs. In this case, the separation between $\chi_A$ and $\chi_B$ is smaller than their characteristic length scale $\xi$. Second, we have the ``true'' Majorana zero modes. In this case,  $\chi_A$ and $\chi_B$ are localized at the opposite ends of the wire and $\phi_\epsilon$ represents the wave function of the (non-local) fermionic mode $\psi^{\dagger} = (\gamma_A+i\gamma_B)/2$  corresponding to the pair of MZMs. Finally, we have the ps-ABSs, which are characterized by component MBSs  (i.e., $\chi_A$ and $\chi_B$) separated by distances comparable with or larger than $\xi$, but less than the length of the wire. The ps-ABSs interpolate continuously between the first two classes and cannot be distinguished  locally from true MZMs separated by the entire length of the wire. In fact, they may also be viewed as MZMs realized in a certain segment of the wire that hosts a local ``topological'' SC phase, while the rest of the wire is topologically trivial. Here, the term ``topological'' is, of course, defined operationally in the context of a finite system -- the finite segment of the wire. Such operationally defined local ``topological'' region with zero energy modes in the topologically trivial phase was found earlier in studies of the energy spectrum of a vortex core in a two-dimensional SM-SC heterostructure with Rashba spin-orbit coupling and Zeeman field proximity induced from a ferromagnetic insulator. \cite{Black} It is important to reiterate, however, that the ps-ABSs occur in the topologically trivial phase, before the topological quantum phase transition indicated by the minimum of the bulk gap, and cannot be used for experimental tests of non-Abelian statistics and topological quantum computation.
We also emphasize that viewing the ps-ABS as a pair of MZMs at the ends of a ``topological segment'' (i.e., a finite region satisfying locally the so-called topological condition \cite{Sau,Annals,Alicea,Long-PRB,Roman,Oreg,Stanescu} $\Gamma^2 > \mu^2 + \Delta_{ind}^2$) could be, strictly speaking, incorrect. More specifically, let us  consider a long wire with a non-uniform electrochemical potential characterized by a constant slope $s = dV_{eff}/dx$. Applying a Zeeman field $\Gamma^2 = (s \ell)^2 + \Delta_{ind}^2$ would satisfy the topological condition in a segment of length $2\ell$. However, the system has an {\em intrinsic} finite length $L_s = \Delta/s$.  As a result, the bulk gap never closes (regardless of the wire length) and the inhomogeneous system is always topologically trivial. Consequently, describing a segment of a wire characterized by an effective potential with finite slope as being in a local ``topological'' phase  is rather ambiguous. In addition, from a more practical point of view, there is a significant difference between the energy splitting of the MZMs in a finite topological wire and the energy splitting of a ps-ABS. More specifically, the energy splitting  characterizing a typical ps-ABS (see, for instance, the ps-ABS modes (IV) and (V) in Fig.~2 and Fig.~4) is much smaller than the splitting of MZMs in a finite wire satisfying the topological condition and with length equal to the separation between the partially overlapping MBSs representing the ps-ABS.

\section{Partially separated ABS in SM-SC heterostructure coupled to a quantum dot}
\label{quantum-dot}
To illustrate the concepts discussed above, let us first focus on a hybrid system consisting of a proximitized nanowire coupled to a quantum dot\cite{Hansen,Prada,Liu-Sau}. The structure is represented schematically in Fig. \ref{XFig1}, together with the corresponding spatial profiles of the induced pairing and effective potential. We note that typically the quantum dot is modeled as a potential valley,\cite{Prada,Liu-Sau} similar to the profile $V_1(x)$ in Fig. \ref{XFig1}(B). However, modeling it as a potential step (see  Fig. \ref{XFig1}(C)) represents another realistic possibility. Which one of the profiles $V_1(x)$ or $V_2(x)$ is a better approximation of the potential characterizing hybrid devices realized in the laboratory depends on the specific materials, in particular the work function difference between the semiconductor and the superconductor, and on the voltage applied to the back gate under the dot region. We find that the assumption of a potential step in the dot region (Fig. \ref{XFig1}(C)) facilitates the creation of ps-ABSs, as one of the constituent MBSs of a low energy ABS induced by the quantum dot can spatially separate to the part of the SM wire in contact with the SC. By contrast, if the potential in the quantum dot is modeled as a valley (Fig. \ref{XFig1}(B)), the constituent MBSs are constrained to live inside the quantum dot as a result of the confining repulsive potential in the part of the SM wire in contact with the superconductor.

\begin{figure}[t]
\begin{center}
\includegraphics[width=0.48\textwidth]{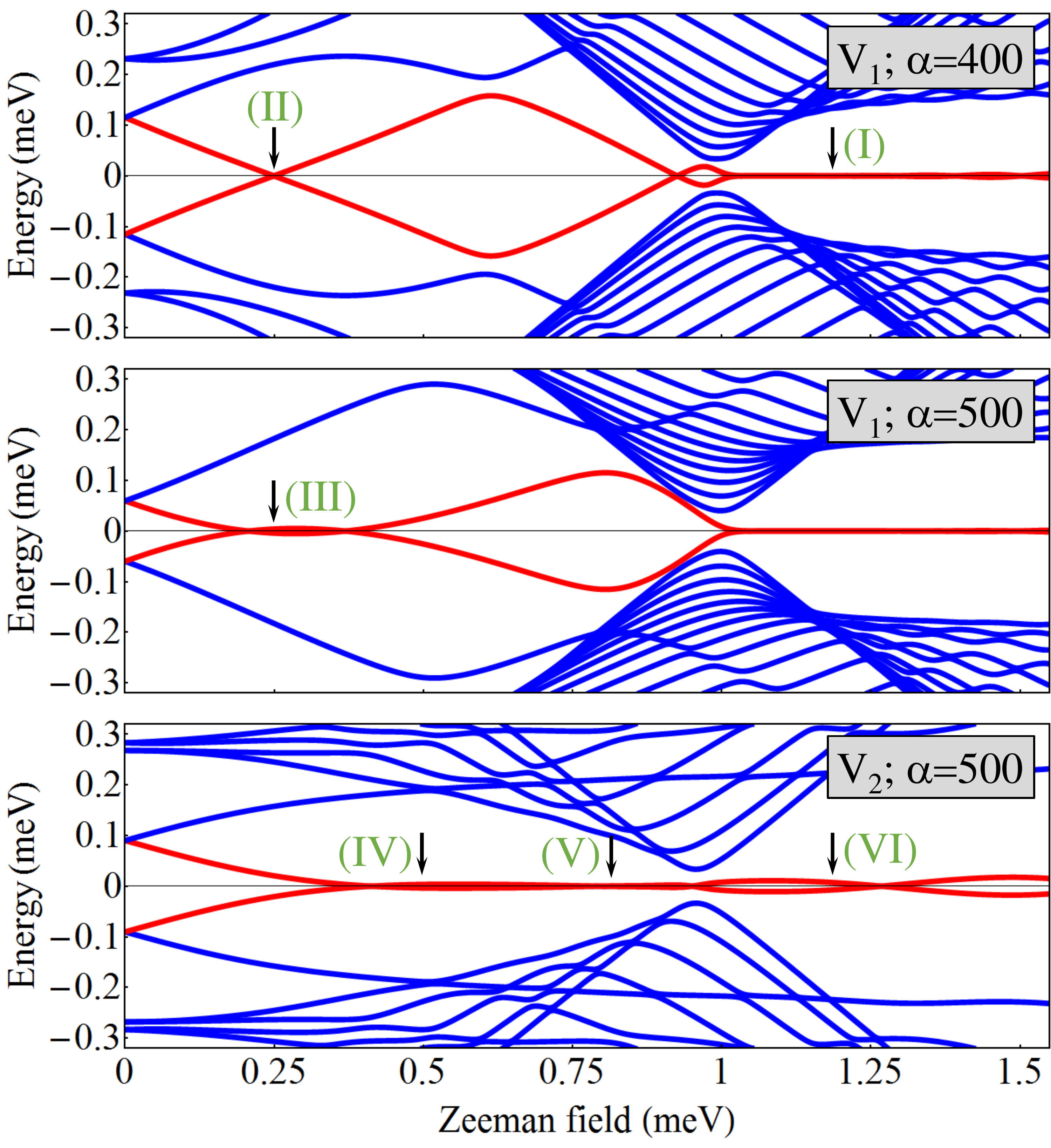}
\end{center}
\vspace{-2mm}
\caption{(Color online) Low-energy spectrum as function of the applied Zeeman field for a system with an effective potential profile $V_1(x)$ (top and middle panels) and $V_2(x)$ (bottom panel). The potential profiles, the system parameters $\Delta_{ind}$ and $\mu$ (measured relative to the bottom of the band), as well as the length of the wire are given in Fig. \ref{XFig1}. The values of the Rashba spin-orbit coupling $\alpha$ indicated on the figure are in units of meV$\cdot$\AA. We have considered a single chain with lattice constant $a=10~$nm, $t_x = 12.7~$meV (corresponding to an effective mass $m_{eff}=0.03m_0$), $t_y=\alpha_y=0$ and the Zeeman field oriented along the wire. The spatial profiles of the MBSs $\chi_A$ and $\chi_B$ associated with the low-energy modes labeled by roman numerals are shown in Figs. \ref{XFig3} and \ref{XFig4}.}
\label{XFig2}
\vspace{-3mm}
\end{figure}

Next, we assume that the chemical potential, which is measured relative to the bottom of the highest energy occupied band, is tuned (using back gates) close to the value of the effective potential in the dot region, as shown in  Fig. \ref{XFig1}, while an external  Zeeman field is applied along the wire. The dependence of the low-energy spectra on the applied field for three different parameter configurations are shown in Fig. \ref{XFig2}. First, we note that
the system undergoes a topological quantum phase transition at a critical field of about $1~$meV, as signaled by the minimun in the bulk spectrum. Of course, the bulk gap remains finite at the ``transition'' as a result of finite size effects. Above the critical field, the system is in a topological SC phase and a ``true'' Majorana zero mode emerges at each end as a mid-gap state. Again, the small but finite energy splitting, which is evident in the lower panel, is the result of finite size effects. Second, we note that a low-energy mode is present even in the trivial regime, i.e., below the ``critical'' field for the topological transition signaled by the bulk gap minimum. In the upper panel of  Fig. \ref{XFig2}, the trivial low-energy mode produces a zero-energy crossing indicative of a ``standard'' ABS. When measured experimentally, such a mode produces a ZBCP that can be easily distinguished from the ZBCP generated by a MZM. However, tuning the system parameters (e.g., the spin-orbit coupling) may result is a situation when the ABSs appear to coalesce at zero-energy, as shown in the middle panel  of  Fig. \ref{XFig2}. The spectroscopic signature of this low-energy mode can still be distinguished from the signature of a MZM if one has experimental access to a wide range of Zeeman fields that extends above the critical field. If, on the other hand, the SC bulk gap collapses before the ``phase transition'', identifying the ZBCP generated by this mode as a signature of a trivial  low-energy state may be more difficult.  In particular, it is necessary to perform consistency tests to check the robustness of the ZBCP (including its quantization at low temperature)  to variations of the controllable experimental  parameters.

While the trivial low-energy mode shown in the middle panel of Fig. \ref{XFig2} is a bad impersonator of a MZM, the scenario illustrated in the lower panel should raise serious concerns regarding the interpretation of the tunneling conductance data \cite{Mourik_2012,Deng_2012,Das_2012,Churchill_2013,Finck_2013,Marcus,Hansen,HZhang,Frolov,Nichele,Quantized-ZBCP}. In this case, the trivial low-energy mode sticks to zero over a wide range of Zeeman fields and is equally robust against the variations of other controllable parameters. In fact, it is virtually impossible to differentiate this mode from a ``true'' MZM using local measurements at the end of the wire containing the quantum dot.

\begin{figure}[t]
\begin{center}
\includegraphics[width=0.48\textwidth]{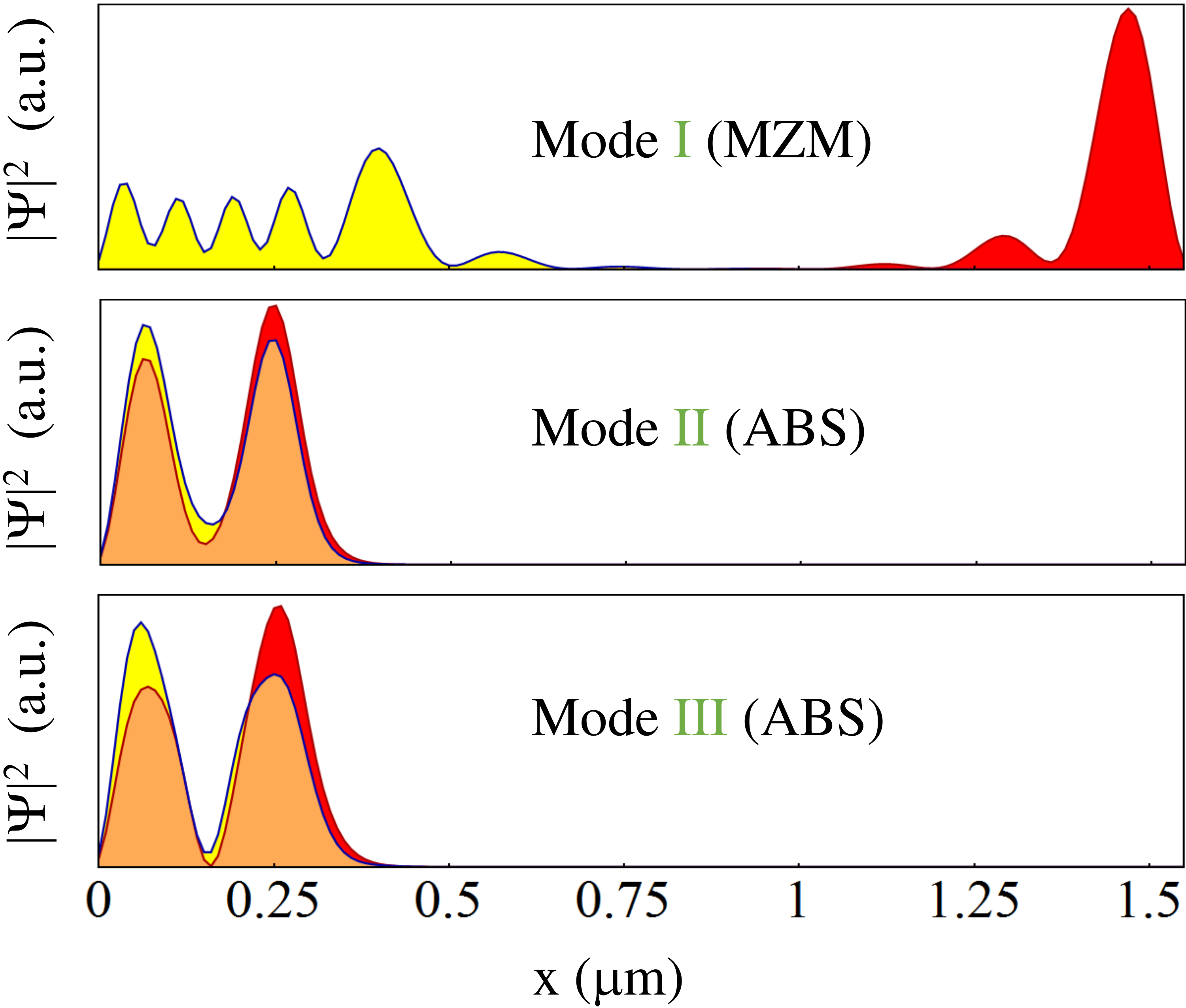}
\end{center}
\vspace{-2mm}
\caption{(Color online)  Spatial profiles of the Majorana bound states $\chi_A$ and $\chi_B$ (yellow/light gray and red/gray, respectively) defined by Eqs. (\ref{chiA}) and (\ref{chiB}) for the low-energy modes labeled by roman numerals in Fig. \ref{XFig2}. Mode (I) is a ``true'' Majorana mode  characterized by two MBSs localized near the opposite ends of the wire (top panel). Modes (II) and (III) are ``standard'' ABSs consisting of two MBSs that are practically on top of each other (middle and lower panels). The values of the model parameters are given in Fig.~\ref{XFig1} and Fig.~\ref{XFig2}.}
\label{XFig3}
\vspace{-3mm}
\end{figure}

To gain more physical intuition, we ``decompose'' the low-energy modes into the constituent MBSs $\chi_A$ and $\chi_B$ as discussed above (see Eq.~\ref{chiA} and Eq.~\ref{chiB}), and analyze their spatial profiles.  Consider first the Majorana mode (I) corresponding to a value of the Zeeman field $\Gamma=1.2~$meV (see Fig. \ref{XFig2}, top panel). As shown in Fig. \ref{XFig3} (top panel), the two MZMs $\chi_A$ and $\chi_B$ are localized at the opposite ends of the wire and have an exponentially small overlap. Note that the MZM localized near the left end of the wire penetrates into the quantum dot, i.e., the normal section of the wire.\cite{Klinovaja-Loss}  Also note that the identification of the yellow/light gray and red/gray MBSs with $\chi_A$ and $\chi_B$, respectively, depends on an overall phase difference between $\phi_\epsilon$ and $\phi_{-\epsilon}$. Introducing an additional $\pi$ phase difference (i.e. a minus sign in one of the wave functions) will switch the positions of  $\chi_A$ and $\chi_B$.

The ABS modes (II) and (III) (see Figs. \ref{XFig2} and \ref{XFig3})) consist of  MBSs that are practically on top of each other. When probing these modes from the left end of the wire one practically couples to both MBSs with similar coupling strengths and the resulting zero bias conductance peaks are (generally) not  quantized. An accidental quantization of the ZBCP can  be identified by varying the control parameters, e.g., the tunnel barrier height.\cite{Liu-Sau}

\begin{figure}[t]
\begin{center}
\includegraphics[width=0.48\textwidth]{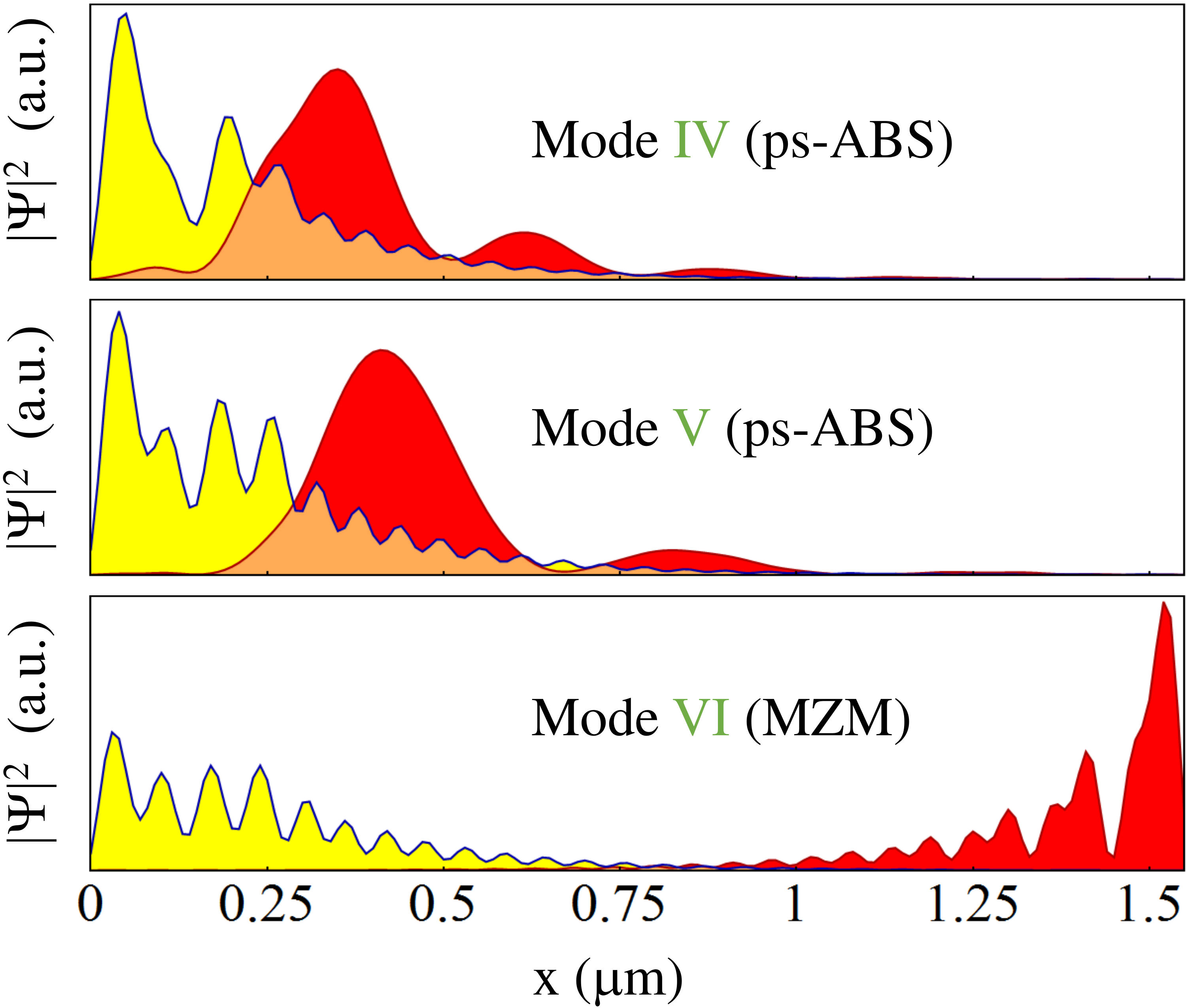}
\end{center}
\vspace{-2mm}
\caption{(Color online)  Spatial profiles of the Majorana bound states $\chi_A$ and $\chi_B$ (yellow/light gray and red/gray, respectively) defined by Eqs. (\ref{chiA}) and (\ref{chiB}) for the low-energy modes marked by roman numerals in the lower panel of Fig. \ref{XFig2}. The trivial low-energy modes (IV) and (V) are partially separated ABSs localized near the left  end of the system, which contains the quantum dot. Mode (VI) consists of two MZMs localized at the opposite ends of the wire.  The values of the model parameters are given in Fig.~\ref{XFig1} and Fig.~\ref{XFig2}.}
\label{XFig4}
\vspace{-3mm}
\end{figure}

The situation is completely different  in the case of the trivial low-energy modes (IV) and (V) (see the lower panel of Fig. \ref{XFig2}) characterizing the system with a step-like effective potential ($V_2$). The spatial profiles of the corresponding MBSs are shown in Fig. \ref{XFig4}. These modes are proper ps-ABSs, as the constituent MBSs are separated by a distance comparable to their length scale (which is of the order $\xi \sim 0.3~\mu$m). In a charge tunneling experiment that probes the left end of the system, the lead will only effectively couple to $\chi_A$, i.e. the yellow/light gray Majorana, while the coupling to the second (red/gray) Majorana will be exponentially small. As a result, the signatures of the trivial low-energy modes (IV) and (V) will be indistinguishable from those of the ``true'' MZM (VI), since the corresponding yellow Majoranas ($\chi_A$) look qualitatively the same at the left end of the wire, while their red counterparts  ($\chi_B$) are practically ``invisible'' to any local probe.

We conclude that the theoretical description of the low-energy modes that occur in semiconductor-superconductor structures in terms of partially separated Andreev bound states represents a powerful tool that allows us to make a direct connection with the spectroscopic features observed experimentally. Using this tool, we are able to make a distinction between bad impersonators of MZMs, which are standard, fine tunned ABSs such as mode (III) described above, and truly worrisome trivial low-energy modes, which are ps-ABSs similar to modes (IV) and (V).

\section{Partially separated ABS with long length scale potential inhomogeneity}
\label{Inhomogeneity}

\begin{figure}[t]
	\begin{center}
		\includegraphics[width=75mm]{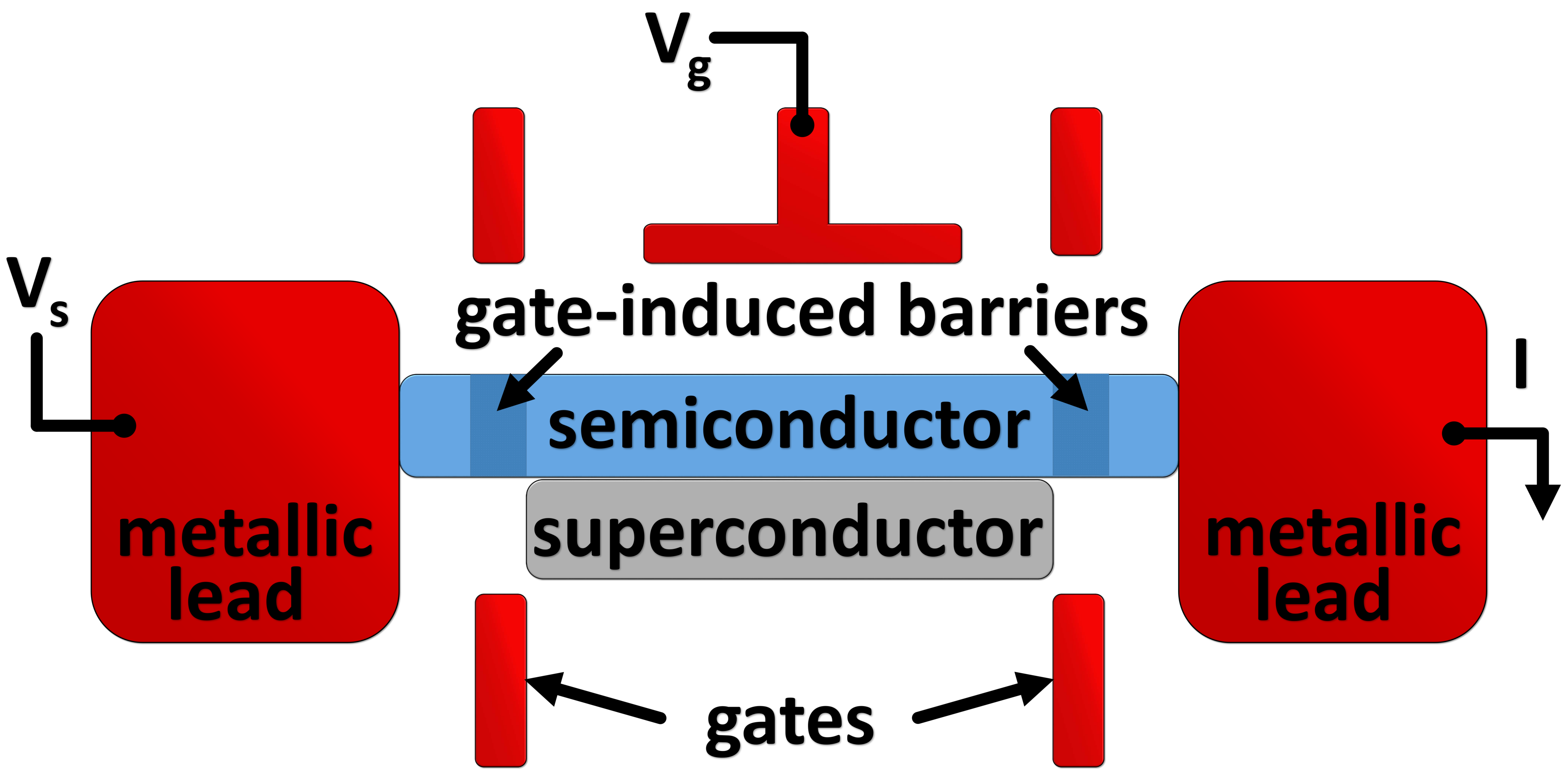}
	\end{center}
	\caption{(Color online) Schematic representation of the experimental set-up in Ref.~[\onlinecite{Marcus}].  Metallic leads are coupled to each end of a semiconductor nanowire proximity-coupled to an s-wave superconductor. Potential gates create tunneling barriers and control the electrostatic energy of the heterostructure. A bias voltage $V_s$ is applied across the wire.} 
	\label{fig:smothConfine}
\end{figure}

Next, we consider a quasi-one-dimensional SM-SC heterostructure with a long length scale potential inhomogeneity in the proximitized nanowire. Such inhomogeneity may arise from the difference in the work functions between the superconductor and the semiconductor, as well as due to the multiple potential gates used in the device (Fig.~\ref{fig:smothConfine}). The long length scale potential inhomogeneities are expected to survive the effects of electron-electron interactions and screening, at least for low occupancy (1-3 occupied bands).\cite{Tudor-tobepublished}

\begin{figure}[t]
	\begin{center}
		\includegraphics[width=70.5mm]{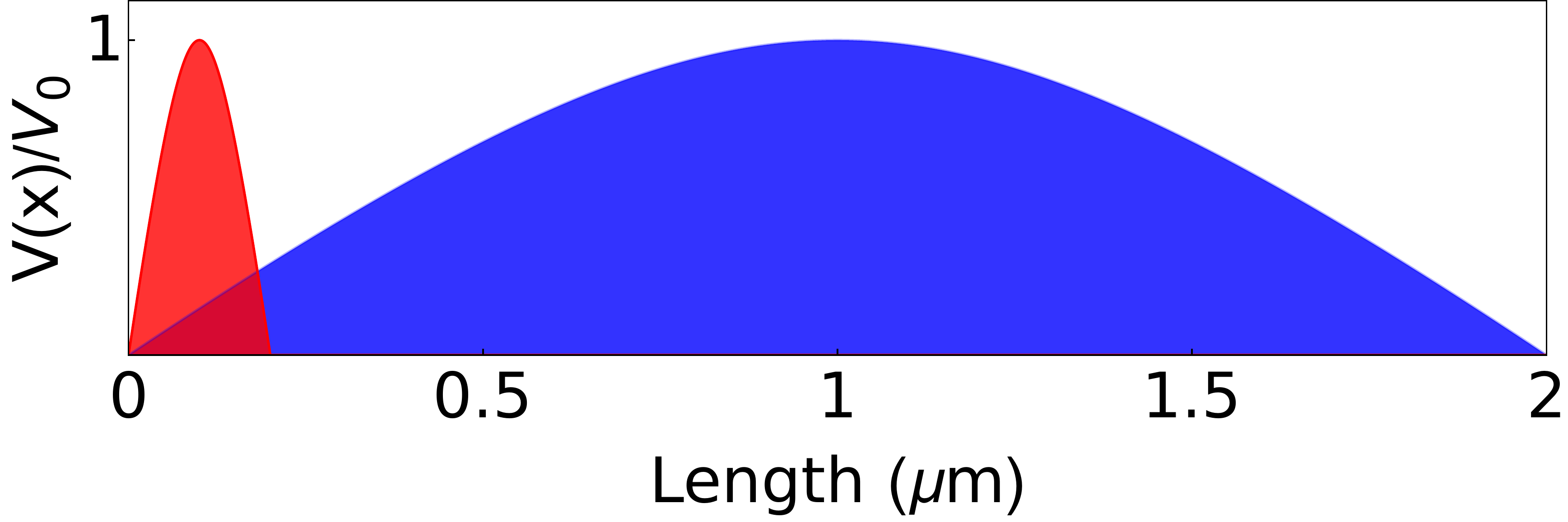}\hspace{-5.5mm}\llap{
			\parbox[b]{139mm}{\large\textbf{(A)}\\\rule{0ex}{19mm}}}		
		\includegraphics[width=75mm]{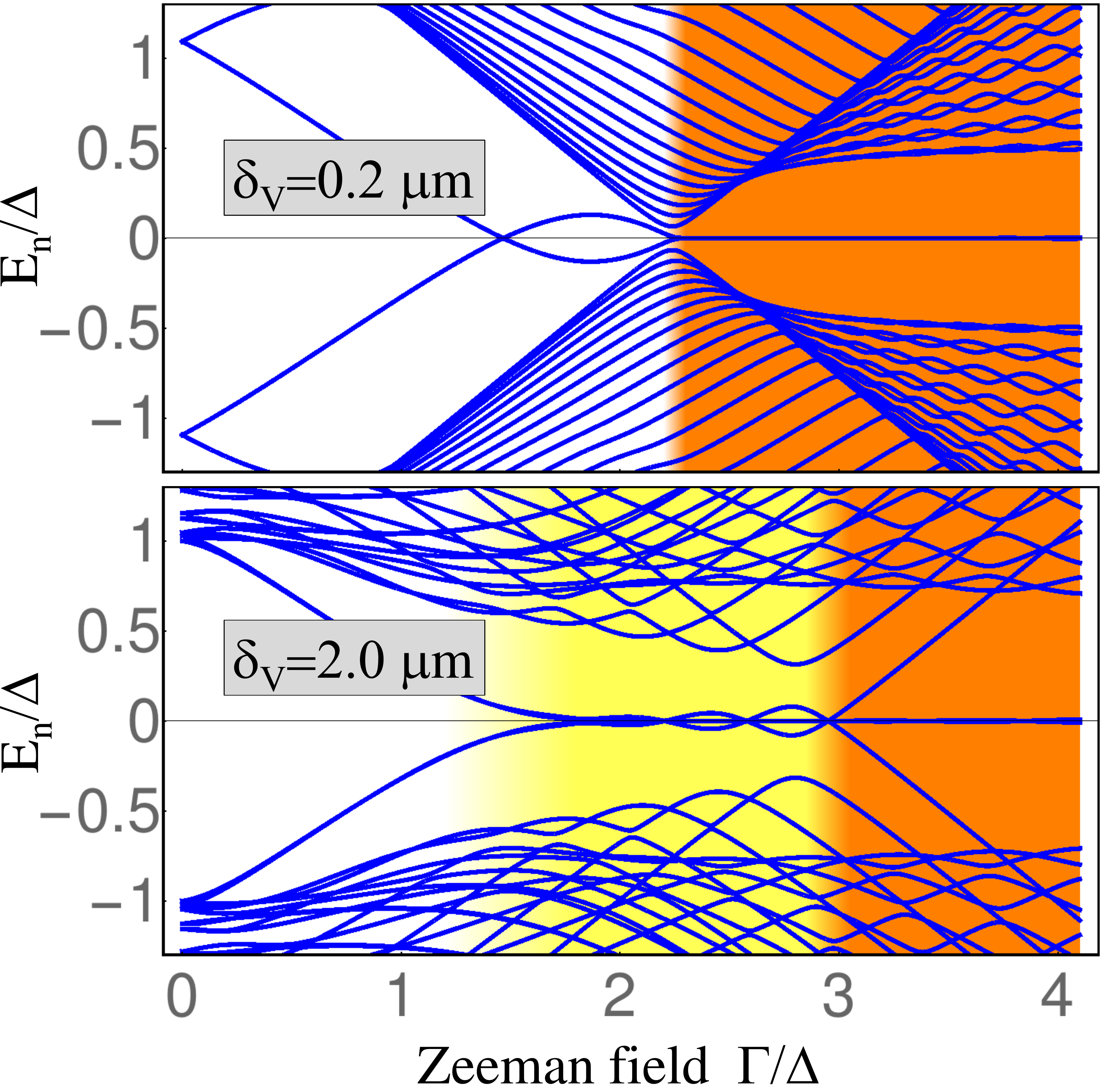}\llap{
			\parbox[b]{149mm}{\large\textbf{(B)}\\\rule{0ex}{30mm}
				\large\textbf{(C)}\\\rule{0ex}{35mm}}}				
	\end{center}
	\caption{(Color online) (A) Inhomogeneous potential profiles with characteristic widths $\delta_V = 0.2~ \mu$m [red (dark gray)] and $\delta_V =2.0~\mu$m [blue (gray)]. The maximum height/depth of the potential  is $V_0$, which can be positive or negative. (B) and (C)  Zeeman field dependence of the low-energy spectra.   (B)  A short-range potential inhomogeneity with $\delta_V = 0.2~ \mu$m and $V_0=-0.55~$meV induces an ABS that crosses zero energy in the topologically-trivial region. A robust MZM emerges in the topological regime [orange (gray)]. (C) A long-range inhomogeneity with $\delta_V = 2~ \mu$m and $V_0=-1~$meV generates four nearly-zero energy MBSs (a ps-ABS near each end) in the topologically-trivial regime [yellow (light gray)], see Fig. \ref{fig:wfPot}. The model parameters are $t_x = 12.7~$meV, $t_y=\alpha_y=0$ (single chain), $\alpha=250~$meV$\cdot$\AA, $\mu=-0.5~$meV [panel(B)], $\mu=-0.75~$meV [panel(C)], $\Delta_{ind}=0.25~$meV, and the length of the wire is $2~\mu$m.}
	\label{fig:FFig4}
\end{figure}

\begin{figure}[t]
	\begin{center}
		\includegraphics[width=75mm]{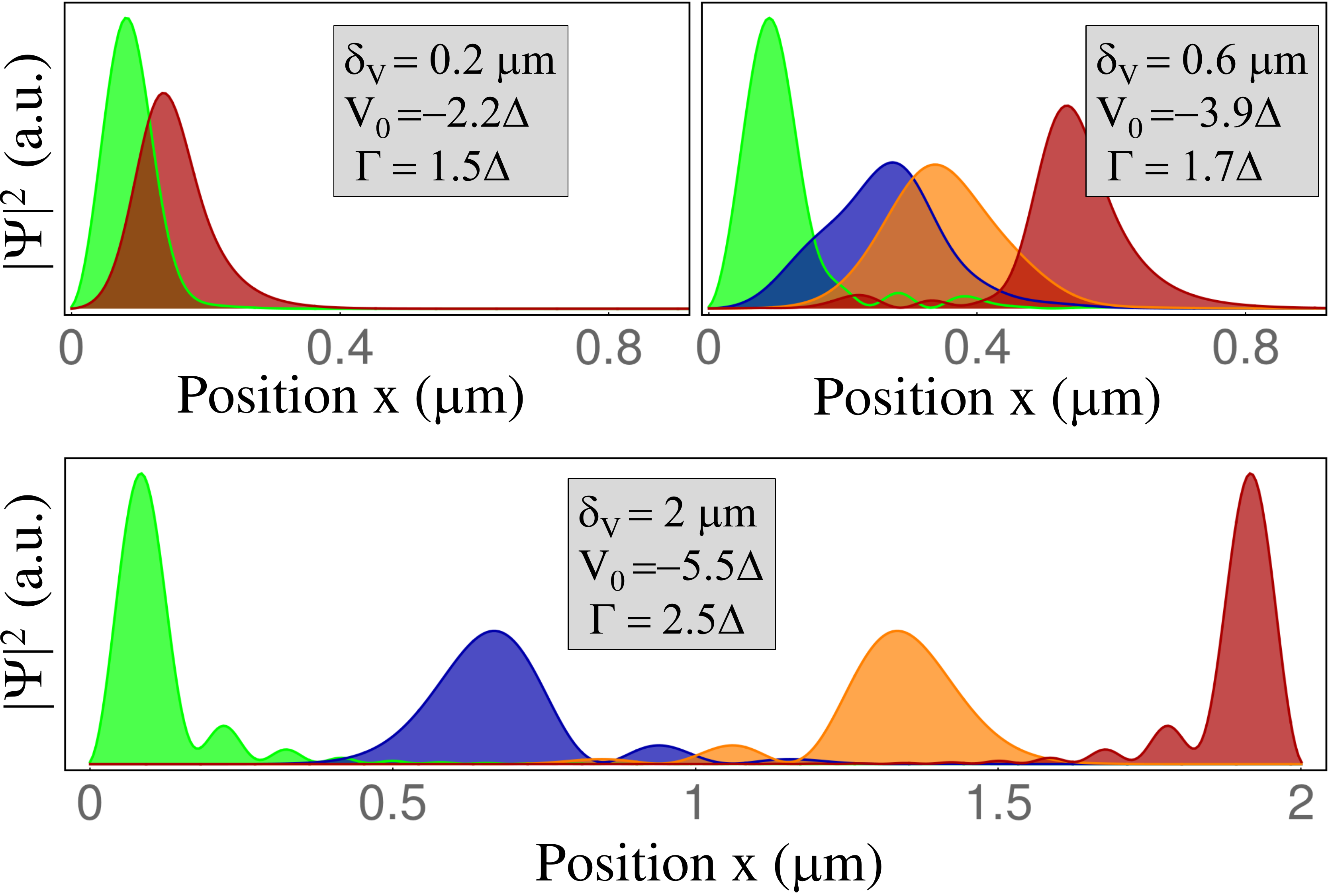}\llap{
			\parbox[b]{130mm}{\large\textbf{(A)}\\\rule{0ex}{25mm}
				\large\textbf{(C)}\\\rule{0ex}{19mm}}}\llap{
			\parbox[b]{58mm}{\large\textbf{(B)}\\\rule{0ex}{45mm}
		}}
		
	\end{center}
	\caption{(Color online) Majorana wave functions $\chi_A, \chi_B$ defined in Eq.~\ref{chiA} and Eq.~\ref{chiB} for the  lowest energy modes in the topologically trivial regime. (A) A localized potential inhomogeneity ($\delta_V = 0.2~ \mu$m) generates two strongly overlapping MBSs, which correspond to a regular ``garden variety'' ABS. Such an Andreev bound state resonance can be easily destabilized from zero energy by a change in the Zeeman field (see Fig.~\ref{fig:FFig4}) (B) By increasing the width of the inhomogeneity ($\delta_V = 0.6~ \mu$m), the lowest energy MBSs $\chi_A,\chi_B$ become spatially separated and another pair of MBSs starts to separate spatially. (C) Four weakly overlapping MBSs forming a pair of ps-ABSs in a system with long-range potential inhomogeneity ($\delta_V = 2~ \mu$m). The green and red states at the two ends originate from the lowest energy spin sub-band, while the blue and orange states originate from the first higher energy spin sub-band. The values of the model parameters are given in Fig.~\ref{fig:FFig4}}
	\label{fig:wfPot}
\end{figure}

In non-homogeneous SM-SC hybrid structures short length scale potential inhomogeneities can generically give rise to zero energy localized Andreev bound states (ABS). \cite{Sau_Demler} While Ref.~[\onlinecite{Sau_Demler}] considered purely local potentials and found zero energy Andreev resonances only in the topological phase ($\Gamma > \Gamma_c$), we find such resonances for short length scale potentials on both sides of the transition, independent of whether the system is topological or trivial.
While for a short length scale inhomogeneity the zero energy resonances are unstable to changes in the Zeeman field (and require fine tuning), for longer length scale potentials they cross over to sub-gap zero energy resonances that are surprisingly robust to perturbations (Fig. \ref{fig:FFig4}), even if the system is topologically trivial.
As in the case considered in the last section, for long length scale potential inhomogeneities inside the proximitized nanowire, it is virtually impossible to distinguish the zero energy robust but trivial ABSs (which are again ps-ABSs, see Fig.~\ref{fig:wfPot}) from spatially well-separated MZMs localized near the ends in the topological phase
using any local measurements.

The low-energy physics of the SM-SC heterostructure in the presence of potential inhomogeneity is investigated using
 a BdG Hamiltonian of the form,
 \begin{equation}
 H=H_0 + \sum_{i,\sigma}V_c\left(i\right)c_{i\sigma}^{\dagger}c_{i\sigma},
  \label{H-Inhomogeneity}
 \end{equation}
 where the inhomogeneous background potential is described by the position-dependent function $V_c(i)$. Typical potential profiles used in the  calculation are shown in Fig. \ref{fig:FFig4} (A).
The low-energy spectrum is obtained by numerically diagonalizing the BdG Hamiltonian. In addition, in the next section, we calculate the tunneling differential conductance in the single-lead and two-lead configurations. Note that in the single-lead configuration the current $I$ is extracted through the SC, while in the two-lead setup the SC is either grounded or isolated, the last case corresponding to Fig. \ref{fig:smothConfine}.


The emergence of potential-induced low-energy modes in the topologically-trivial regime is illustrated in Fig. \ref{fig:FFig4}.
First, we consider a short-range potential corresponding to the red (dark gray) curve in Fig.  \ref{fig:FFig4}~(A). The potential induces an Andreev bound state that crosses zero energy in the topologically trivial regime, as shown in panel (B).
 Next, we consider the potential corresponding to the blue (gray) curve in Fig. \ref{fig:FFig4}~(A). In this case, upon increasing the Zeeman field, four low-energy states (a pair of weakly overlapping MBSs at each end) emerge while the system is still in the topologically-trivial regime, as shown in panel (C).

The partially-overlapping MBSs responsible for the low-energy modes discussed above are shown in Fig. \ref{fig:wfPot}. The ABS induced by the localized potential (see Fig. \ref{fig:FFig4}B) corresponds to two strongly overlapping MBSs, as shown in panel \ref{fig:wfPot}A. Increasing the width of the potential results in a larger spatial separation between the two MBSs $\chi_A$ and $\chi_B$. In addition, another pair of low-energy overlapping MBSs  emerges (see  panel \ref{fig:wfPot}B). The overlapping pairs of MBSs (with separation less than the characteristic Majorana decay length $\xi$) in panels A and B can be easily distinguished from topological MZMs (with separation between $\chi_A$ and $\chi_B$ $\sim L$) by examining the stability of the corresponding ZBCPs to various perturbations.  However,  the low-energy bound states populating the yellow (light gray) topologically trivial regime ($\Gamma < \Gamma_c$) in Fig. \ref{fig:FFig4}C are the four weakly overlapping MBSs (a ps-ABS at each end) shown in panel \ref{fig:wfPot}C.

As before, we conclude that while the trivial low-energy modes shown in Figs. \ref{fig:wfPot}A and \ref{fig:wfPot}B) cannot be confused with MZMs because the corresponding ZBCPs will be quickly split by the Zeeman field, barrier potential, and other perturbations, the scenario illustrated in Fig.~\ref{fig:wfPot}C  should be a matter of serious concern for identifying the ZBCPs in the existing tunneling conductance data with topological MZMs \cite{Mourik_2012,Deng_2012,Das_2012,Churchill_2013,Finck_2013,Marcus,Hansen,HZhang,Frolov,Nichele,Quantized-ZBCP}. In this case, not only do the trivial low-energy modes (a ps-ABS at each end) stick to zero energy over a wide range of Zeeman fields, chemical potential, and other external controllable parameters, as we show in the next section it is virtually impossible to differentiate the ps-ABSs from a ``true'' MZM in the topological phase using any conceivable local measurement including the measurement of two-terminal resonant charge transfer in the Coulomb blockade regime \cite{Marcus}.
 In general, if the effective potential in the semiconductor wire has variations of the order of the induced gap on length scales comparable to the Majorana localization length, the low-energy modes will correspond to a ``Majorana chain'' similar to the situation illustrated in Fig. \ref{fig:wfPot}C. Note that these MBSs can be associated with either one of the top occupied spin-split sub-bands.

\begin{figure}
	\includegraphics[width=75mm]{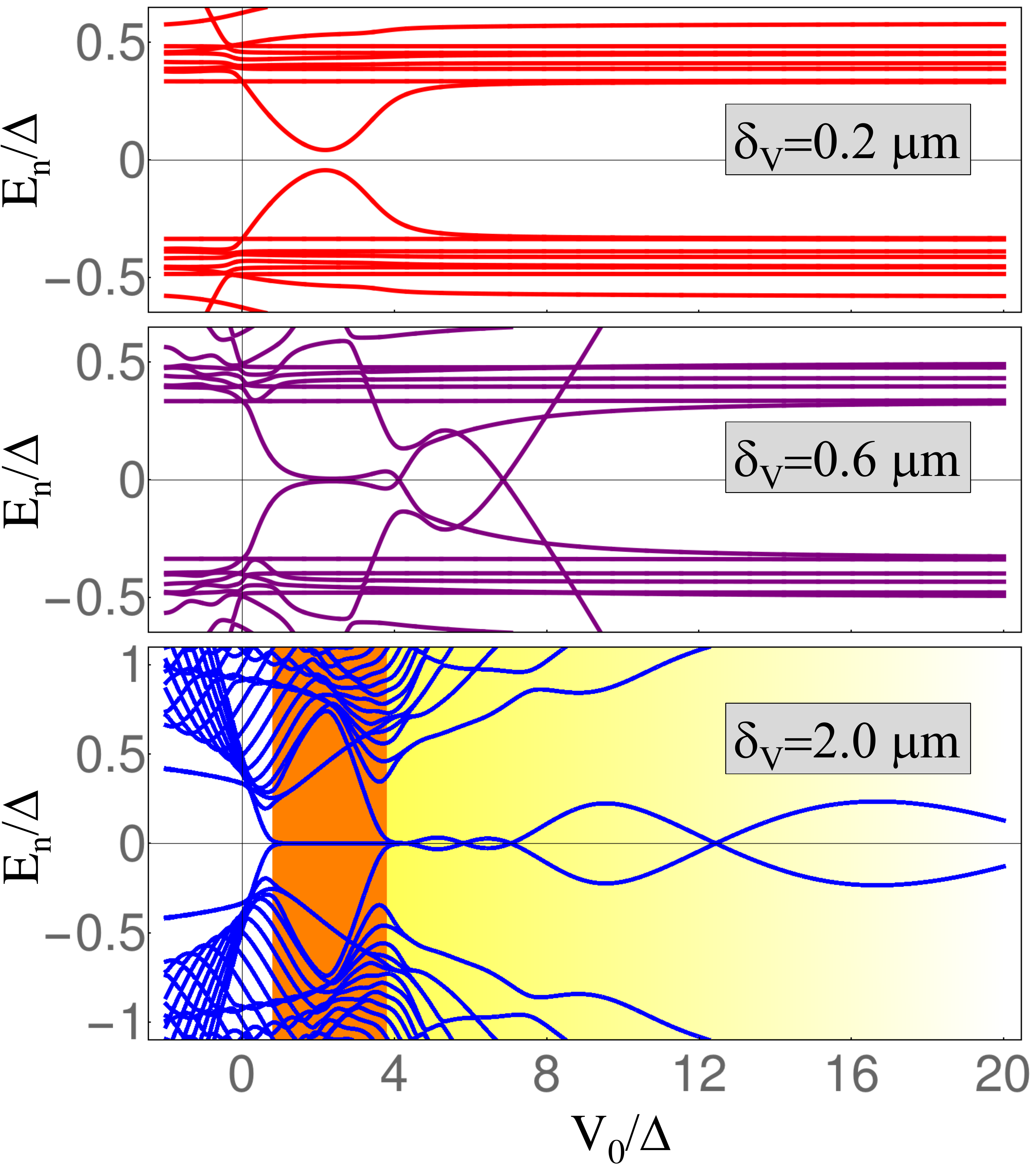}\llap{
		\parbox[b]{155mm}{\large\textbf{(A)}\\\rule{0ex}{23mm}
			\large\textbf{(B)}\\\rule{0ex}{23mm}	\large\textbf{(C)}\\\rule{0ex}{32mm}}}	
	\caption{(Color online) Low-energy spectra in the presence of a non-homogeneous potential. The Zeeman field is $\Gamma=1.75\Delta$ and the chemical potential (defined relative to the bottom of the top band when $V_0=0$) is $\mu=2\Delta$. The values of the other parameters are given in Fig.~\ref{fig:FFig4}. As the characteristic length scale $\delta_V$ increases, robust low-energy ps-ABSs emerge in the topologically-trivial regime [yellow (light gray) regions in panel C].}
	\label{fig:FFig2}	
\end{figure}

The dependence of the low-energy modes associated with the partially overlapping MBSs on the inhomogeneous  potential is shown in Fig.~\ref{fig:FFig2}. For short-range potentials, $\delta_V\ll L$, the parameters of the calculation ($\Gamma=1.75\Delta$ and $|\mu|=2\Delta$) correspond to the topologically trivial regime. Increasing the characteristic length scale $\delta_V$ of the potential inhomogeneity stabilizes these low-energy modes (panel B) as a result of increasing the spatial separation between the MBSs. For large-enough values of $\delta_V$, {\em effectively zero-energy} modes emerge in both the topological (orange/dark gray) and trivial (yellow/light gray) regimes (panel C).


\begin{figure}
	\begin{center}
		\includegraphics[width=80mm]{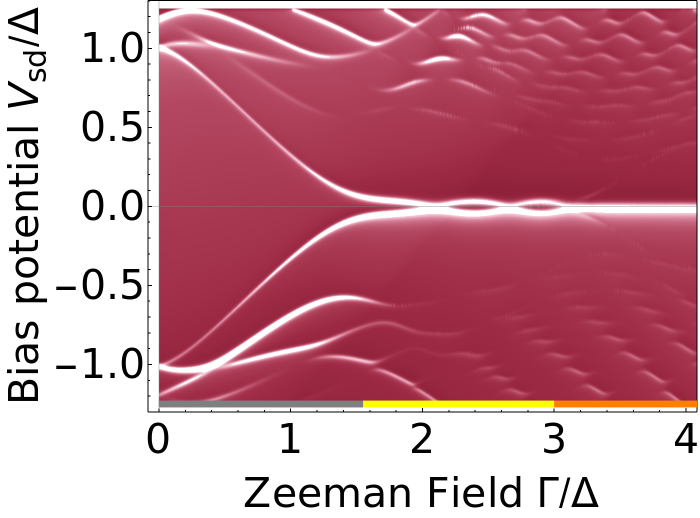}\vspace{3mm}
		\includegraphics[width=80mm]{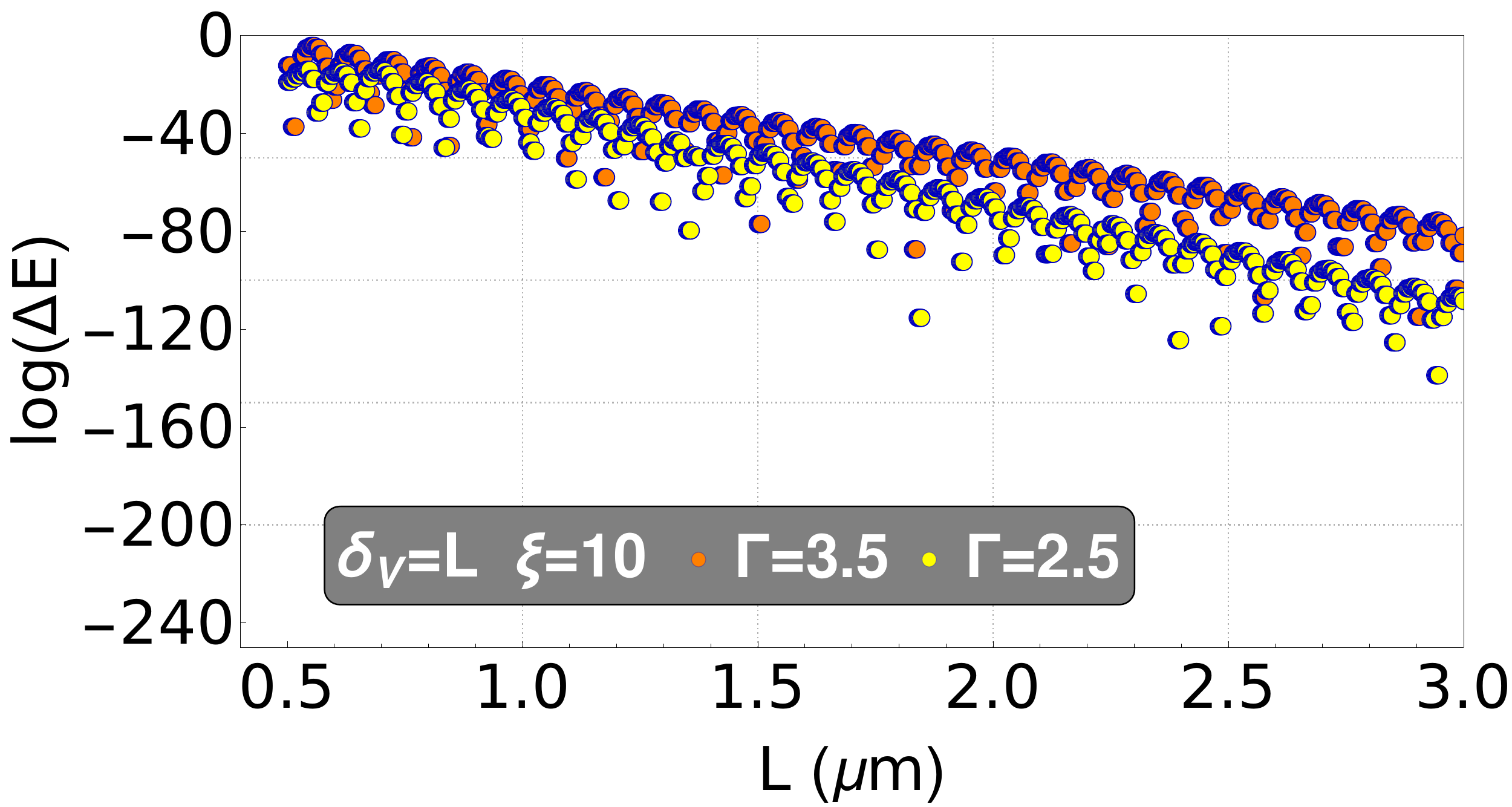}
	\end{center}				
	\caption{(Color online)(top) Measurements of the local density of states (LDOS) as a function of bias potential and Zeeman splitting associated with a long length scale potential inhomogeneity with characteristic length $\delta_{V}=2.0\mu m$ (Fig.~\ref{fig:FFig4}). For sufficiently long potential length scales zero energy crossings in the LDOS spectrum exist in the topologically trivial regime [yellow (light gray)] due to a ps-ABS at each end. (bottom) Logarithm of the energy splitting $\Delta E\propto\exp(-L/\xi)$ as a function of wire length $L$ showing exponential protection of the energy splitting 
in both the topologically trivial regime [yellow (light gray) dots, $\Gamma=2.5$] and the topological regime [orange (gray) dots, $\Gamma=3.5$. LDOS was taken at a temperature of $T\approx20$mK, with the values of the other parameters given in Fig.~\ref{fig:FFig4}. In the trivial regime the energy splitting is exponentially suppressed in the length of the wire only when the length scale of the potential fluctuation is given by the length of the wire $L$.
.}
	\label{fig:LDOS}	
\end{figure}


\begin{figure}
	\begin{center}
		\includegraphics[width=75mm]{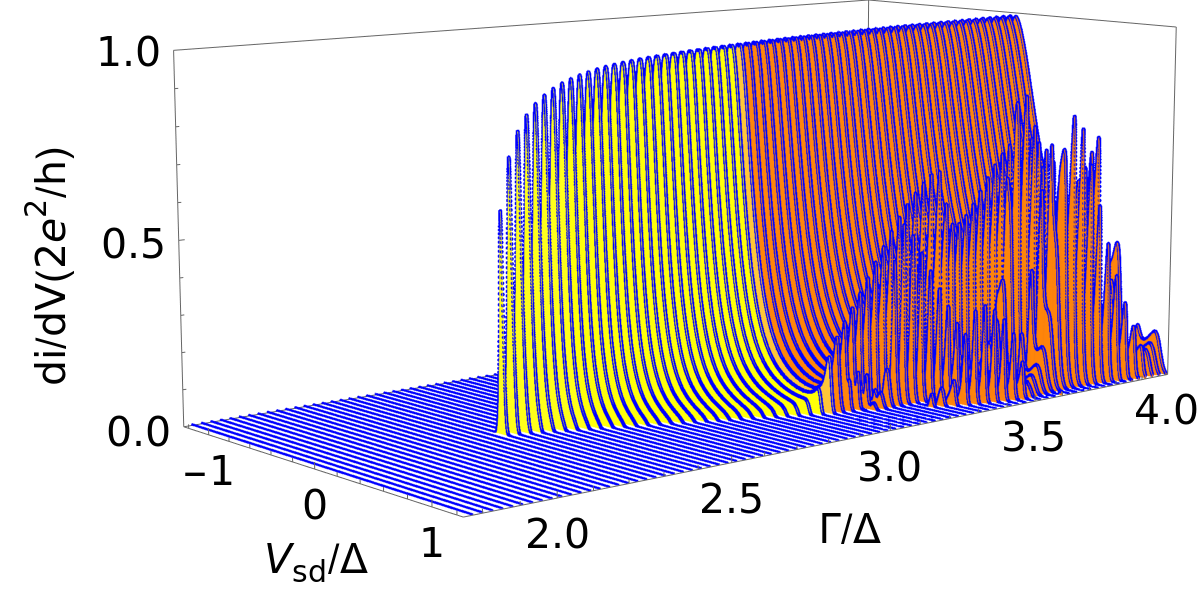}\llap{
			\parbox[b]{118mm}{\large\textbf{(A)}\\\rule{0ex}{28mm}}}
		\includegraphics[width=75mm]{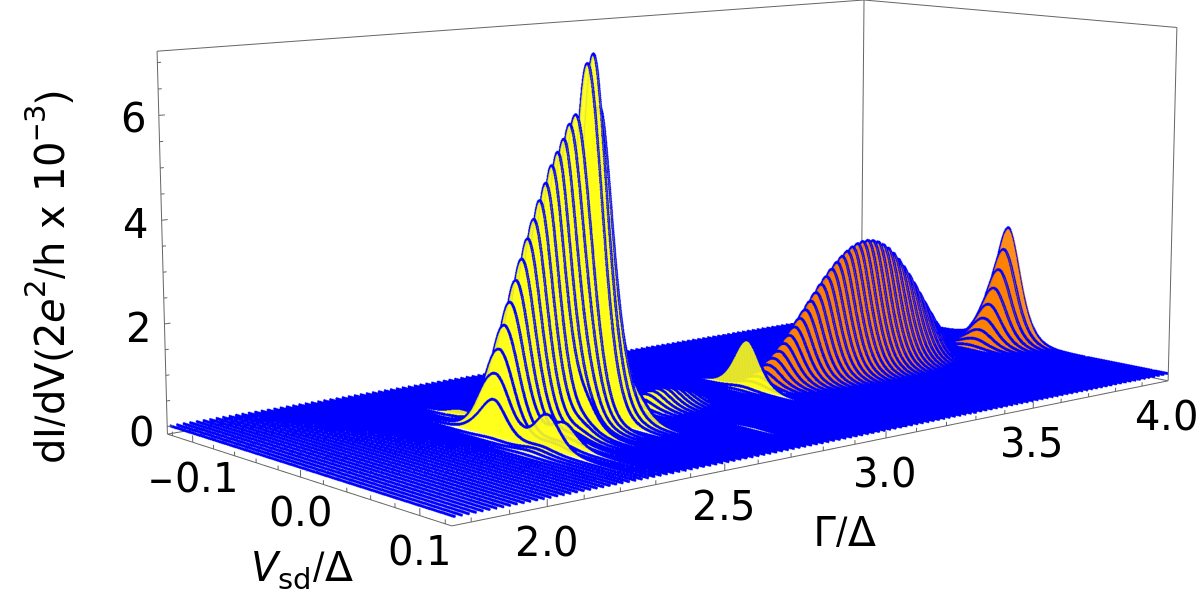}\llap{
			\parbox[b]{118mm}{\large\textbf{(B)}\\\rule{0ex}{28mm}}}
		\includegraphics[width=75mm]{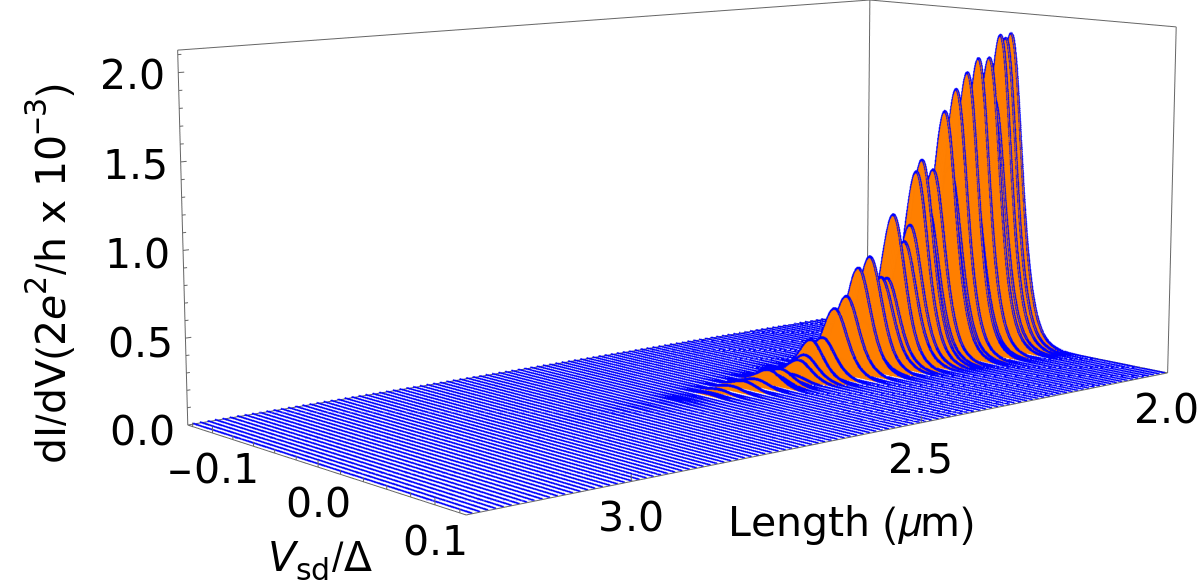}\llap{
			\parbox[b]{118mm}{\large\textbf{(C)}\\\rule{0ex}{28mm}}}	
	\end{center}
	\caption{(Color online)(A) Single lead differential conductance plot of a $2\mu m$ superconductor-semiconductor heterostructure with a normal lead at one end of the wire. The Zeeman field ranges from $\Gamma=1.5\Delta$ to $\Gamma=4\Delta$ with profiles offset for clarity. (B) Two lead differential conductance in the Coulomb blockade regime  as a function of Bias potential $V$ and Zeeman splitting $\Gamma$ with long length scale potential fluctuations $\delta_V=2.0\mu m$ . Peaks exist in both the topologically trivial [yellow (light gray)]  and the topological [orange (gray)] regimes.  These peaks are suppressed at zero energy crossings in the spectrum. (C) Two-lead ``teleportation'' differential conductance as a function of bias potential at $\Gamma = 4\Delta$, showing exponential suppression with length of the wire.  In this figure, all conductance calculations were performed for a chain at a temperature $T\approx20$mK, with the values of the other parameters given in Fig.~\ref{fig:FFig4}
}
	\label{fig:cond}
\end{figure}

\section{Single lead charge tunneling and teleportation}
\label{Probe}
Given the ubiquity of stable low-energy modes associated with ps-ABSs in SM-SC heterostructures with spin-orbit coupling and Zeeman fields, the following key question arises: how can one distinguish experimentally between ps-ABSs, which are robust near-zero energy excitations emerging in the trivial regime ($\Gamma < \Gamma_c$), and spatially well separated, non-degenerate,  MZMs associated with the topological phase $\Gamma > \Gamma_c$? We find that ps-ABSs emerging in the trivial phase are indistinguishable from topological MZMs, if  we use localized conductance measurements.  As shown in Fig. \ref{fig:LDOS}, the local density of states (LDOS) has a pronounced peak at zero energy in both the topological regime (orange) with a single MZM localized near each end, as well as  in the topologically trivial regime (yellow)  with a pair of ps-ABSs  near each end (Fig. \ref{fig:wfPot}C). Note that the energy splittings associated with these modes are exponentially protected with increasing wire length in both the trivial and topological regimes, as shown in the bottom panel. The energy splittings in the topologically trivial phase are exponentially suppressed in the length of the wire only when the length scale of the potential fluctuation is also given by the length of the wire $L$.

In Fig.\ref{fig:cond} (A) we show the differential conductance dI/dV for a single-lead set up similar to that in  Ref. \cite{Mourik_2012}.
The expected signature of MZMs located at the ends of the wire is a zero bias conductance peak, which results from resonant local Andreev tunneling. However,  the zero bias peak actually extends smoothly  into the topologically trivial regime (yellow) with robust ps-ABSs at each end generated by a long length scale potential inhomogeneity.

More recently, experiments on SM-SC heterostructures have been carried out in the Coulomb blockade regime \cite{Marcus}, where charging energy discriminates between states with different numbers of electrons and resonant Andreev tunneling is suppressed,
allowing the coherent transport of a single electron via the complex fermionic mode $\psi^{\dagger}=\gamma_1+i\gamma_2$ composed out of the MZMs localized near the ends. This process,  sometimes called ``teleportation'' \cite{Fu,Sau_Swingle_Tewari,Tewari_Zhang_Sarma}, is observable as a zero bias conductance peak periodic in the gate-induced charge $N_g$ with a period $e$ \cite{Marcus,Heck}.
In Fig. \ref{fig:cond} (B) we show the differential conductance as a function of bias potential and Zeeman field for a two-lead set up (Fig.~\ref{fig:smothConfine}) in the Coulomb blockade regime. We incorporate the effects of the charging energy $E_g$ by suppressing the anomalous tunneling (Andreev) processes at the lead-superconductor interfaces.
In the topological regime,  the remaining MZM-assisted charge tunneling process between the two metallic leads is expected to represent the ``teleportation'' amplitude.
In Fig. \ref{fig:cond} (C), we show the exponential fall-off of the zero bias conductance peak in the Coulomb blockade regime for Zeeman splitting corresponding to the topological regime. This behavior is due to the exponential suppression of the wave function overlap between the MZMs in the topological phase.
However, we find a similar exponential fall off in the topologically trivial regime with robust ps-Andreev states due to potential fluctuations. In Fig. \ref{fig:cond} ( B) , we show that well-defined zero bias peaks resulting from a two-lead set-up actually exist in both the topologically trivial (yellow) and non-trivial (orange) regimes in the presence of ps-ABSs with long length scale potential inhomogeneity in the SM.

\section{Two-terminal charge tunneling: Disentangling MZMs from partially separated ABSs}
\label{Two-terminal}


As seen above, the ps-ABSs are a generic low-energy feature of spin-orbit coupled SM-SC heterostructures in the presence of a (suitably directed) Zeeman field. Local measurements (e.g., one-terminal charge or spin tunneling) cannot distinguish between them and non-Abelian MZMs. The ps-ABSs, on the other hand, cannot be used to demonstrate non-Abelian statistics or to perform topological quantum operations, which require  well separated MZMs localized at the ends of the wire.
Below we show that the simplest and most straightforward way of discriminating between a ps-ABS and a topological MZM (i.e., Majorana zero modes which can be used in topological quantum computation) is to perform separate tunneling measurements at the two ends of the wire and look for correlations. We will show that the absence of correlations is indicative of the presence of ps-ABSs, while the presence of correlations is consistent with a topological MZM.


Our proposed experiment to discriminate between ps-ABSs and topological MZMs involves collecting two sets of data. The first data set records the differential conductance spectra at the left end of the wire by applying a bias potential to the left lead, with the superconductor grounded and the right lead isolated. The second data set consists of the differential conductance spectra at the right end by applying a bias potential to the right lead, with the superconductor again grounded and the left lead isolated. The tunneling potentials (as well as all other gate potentials) are kept fixed throughout the measurement for straightforward comparison between the two sets of data. With ps-ABSs created by non-homogeneous potentials (the case of lead-quantum dot-proximitized nanowire hybrid structure \cite{Hansen,Prada,Liu-Sau} is analogous), we show below that comparing the left and the right differential conductances, $(dI/dV)_L$ and $(dI/dV)_R$, provides essential information about the bulk state of the wire, allowing us to discriminate between ps-ABSs and topological MZMs.


\begin{figure}[t]
\begin{center}
\includegraphics[width=0.48\textwidth]{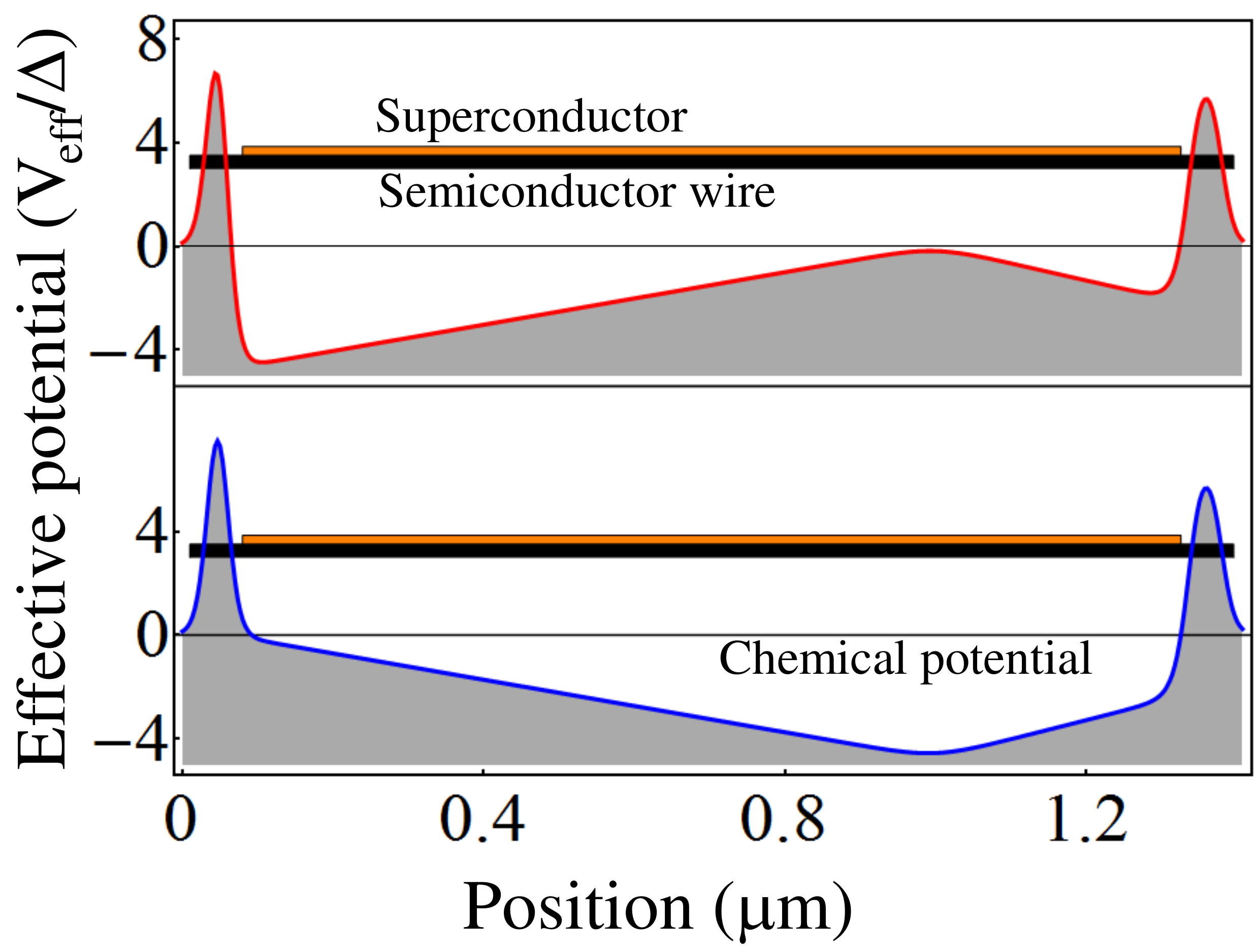}
\end{center}
\vspace{-2mm}
\caption{(Color online) Schematic representation of the inhomogeneous semiconductor-superconductor device along with the two effective potential profiles considered in this study.  The potential maxima at the ends of the wire represent the left and right tunnel barriers. The effective potential is given in units of the induced gap, $\Delta_{ind} = 0.25~$meV, while the chemical potential can be controlled using back gates (here we show the case corresponding to $\mu=0$).}
\label{FIG1}
\vspace{-3mm}
\end{figure}





Real nanowires are 3D systems and the electrostatic potential inside the wire is a position-dependent function $V(x,y,z)$.  We note that, in general, the position dependence of $V$ is determined by the work function difference between the semiconductor and the superconductor,  applied gate potentials, and non-homogeneous charge distributions. By self-consistently solving a three-dimensional Schrodinger-Poisson equation it can be shown that the inhomogeneous potentials in the proximitized nanowire can survive electron-electron interaction and associated screening effects at least in low occupancy wires with 1-3 occupied bands. \cite{Tudor-tobepublished} However, even in the absence of potential inhomogeneities in highly clean ballistic nanowires, the general ideas involving ps-ABSs and utility of two-terminal charge tunneling experiments are more generally valid, as should be clear from the discussion of lead-quantum dot-proximitized nanowire heterostructure \cite{Hansen,Prada,Liu-Sau} in Sec.~\ref{quantum-dot}.

Focusing, for simplicity, on a specific confinement-induced band described (approximately) by the transverse wave-function $\psi_n(y,z)$, we can define a one-dimensional {\em effective potential} as
\begin{equation}
V_{\rm eff}^{(n)}(x) = \langle \psi_n |V(x) |\psi_n\rangle,
\end{equation}
where the matrix element involves a double integral over the transverse coordinates $y,z$.
For the purpose of this paper, we use a simplified 1D model and assume that the effective potential has one of the profiles shown in Fig. \ref{FIG1}. Note that the maximum variation of $V_{\rm eff}$ along the proximitized segment of the wire (i.e., not including the tunnel barriers) is about $1~$meV.
\begin{figure}[t]
\begin{center}
\includegraphics[width=0.48\textwidth]{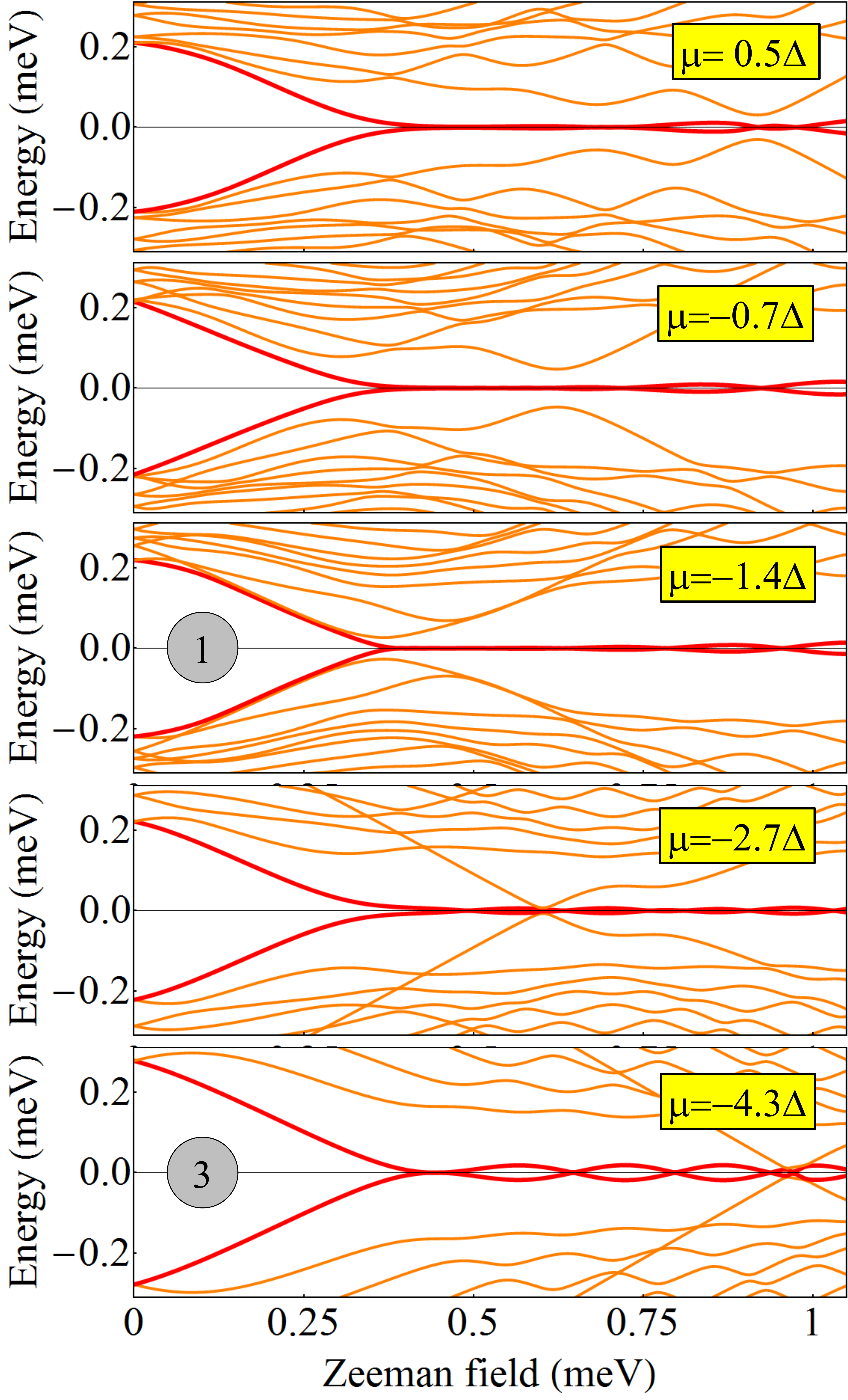}
\end{center}
\vspace{-2mm}
\caption{(Color online) Low-energy spectrum as a function of the applied Zeeman field for the effective potential profile shown in the top panel of Fig. \ref{FIG1} and different values of the chemical potential. The lowest energy modes (red lines)  represent a pair of partially overlapping MBSs that, in general, are not localized near the ends of the wire. The model parameters are $t_x = 12.7~$meV, $t_y=\alpha_y=0$ (single chain), $\alpha=250~$meV$\cdot$\AA, and $\Delta_{ind}=0.25~$meV.}
\label{FIG2}
\vspace{-3mm}
\end{figure}
In Fig~\ref{FIG2} and Fig.~\ref{FIG3} we show the calculated BdG spectra of the proximitized nanowire with the inhomogeneous effective potential profiles given in the top and the bottom panels of Fig.~\ref{FIG1}, respectively. The lowest energy states (red lines in Fig. \ref{FIG2} and blue lines in Fig.~\ref{FIG3}) coalesce toward zero energy and form  robust (nearly) zero-energy modes at Zeeman fields above (approximately) $0.4~$meV.  Note that the ``critical field'' at which the (nearly) zero-energy mode emerges depends weakly on the chemical potential over a range of about $1~$meV. This behavior is clearly inconsistent with having a topological quantum phase transition (and the emergence of topological MZMs separated by the wire length $L$) in a homogeneous system, where, in weak coupling, the ``critical'' Zeeman field $\Gamma_c$ is given by $\Gamma_c = \sqrt{\Delta_{ind}^2 + \mu^2}$, and is thus strongly dependent on the chemical potential. Furthermore, the presence of additional low-energy sub-gap  modes (for certain values of the chemical potential) represents another indication that the system is non-homogeneous. Nonetheless, the lowest energy mode looks extremely robust, based on how strongly it ``sticks'' to zero energy, and is thus for all practical purposes indistinguishable from a topological MZM.  Hence, the key question is, does this mode represent a pair of ``true'' MZMs localized at the ends of the wire or a pair of ps-ABSs in the topologically trivial phase ? To answer this question, we determine the differential conductance for charge tunneling into the two ends of the wire and compare them looking for signs of correlations.

\begin{figure}[t]
\begin{center}
\includegraphics[width=0.48\textwidth]{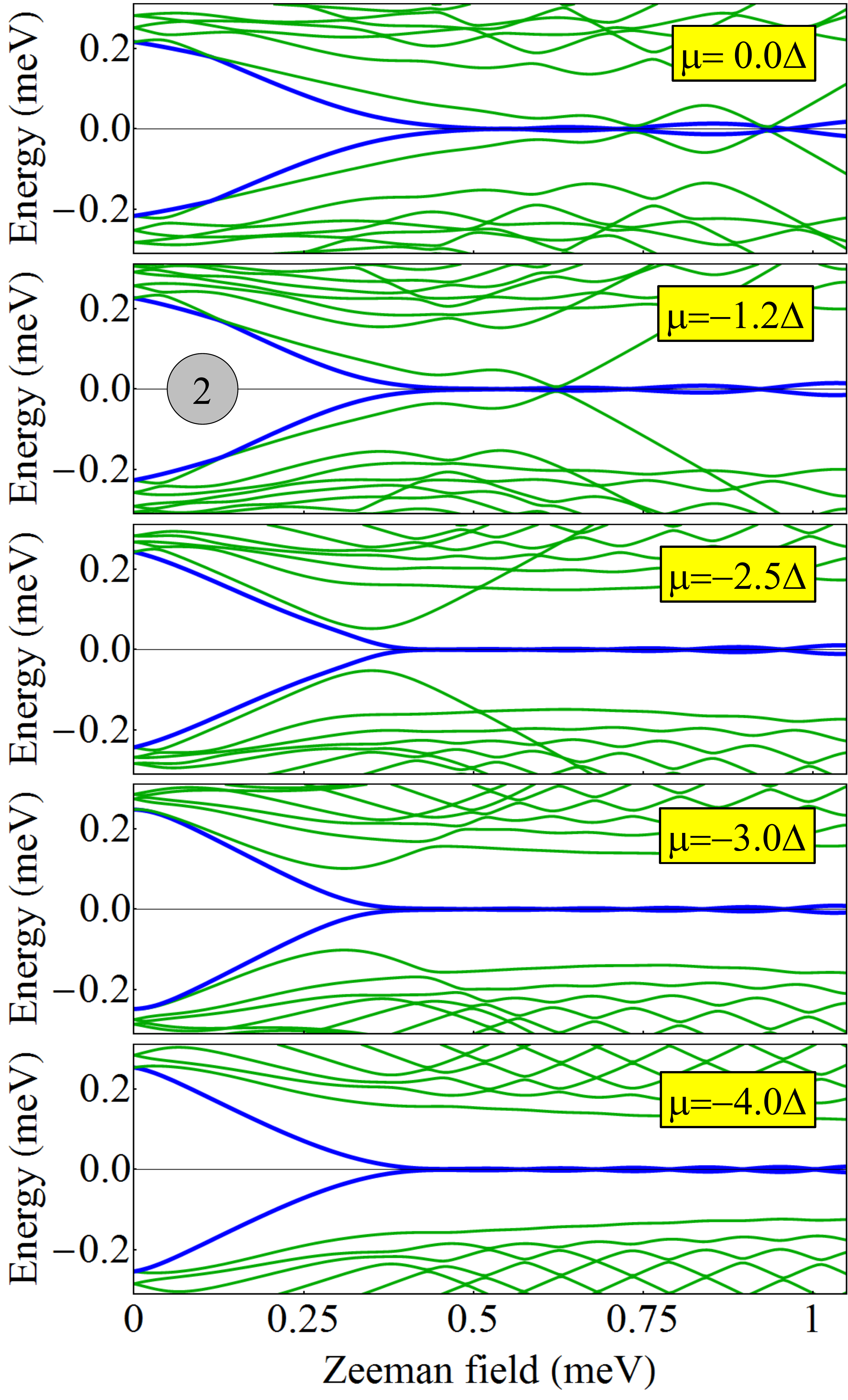}
\end{center}
\vspace{-2mm}
\caption{(Color online) Low-energy spectrum as function of the applied Zeeman field for the effective potential profile shown in the bottom panel of Fig. \ref{FIG1} and different values of the chemical potential. The blue lines represent partially overlapping MBSs. Note that for certain values of the chemical potential there are additional sub-gap low-energy modes. The model parameters are given in Fig. \ref{FIG2}.}
\label{FIG3}
\vspace{-3mm}
\end{figure}

Before we compare the tunneling spectra in the left (L) and the right (R) leads for parameter sets marked by the numbers ``1, 3'', and ``2'' in Fig.~\ref{FIG2} and Fig.~\ref{FIG3}, respectively, let us note that in a non-homogeneous system, the general condition for having MZMs localized  at the ends of the wire is (roughly)  that the maximum of the effective potential variation (relative to the chemical potential) be smaller than the Zeeman splitting (which has to satisfy the topological condition everywhere). In principle, there is always a magnetic field large enough to satisfy this requirement, but this value can be practically irrelevant (e.g., if it leads to the collapse of the bulk SC gap). Nonetheless,  if the ``effectively homogeneous'' condition is realized for Zeeman fields $\Gamma > \Gamma^*$, the L and the R spectra are in general uncorrelated for $\Gamma < \Gamma^*$, while showing correlated features for Zeeman fields larger than $\Gamma=\Gamma^*$. Note that the value of $\Gamma^*$ (which depends on the variation of the effective  potential  relative to the chemical potential) can be changed using a back gate potential that generates an overall shift of $V_{\rm eff}$ with respect to $\mu$.

\begin{figure}[t]
\begin{center}
\includegraphics[width=0.48\textwidth]{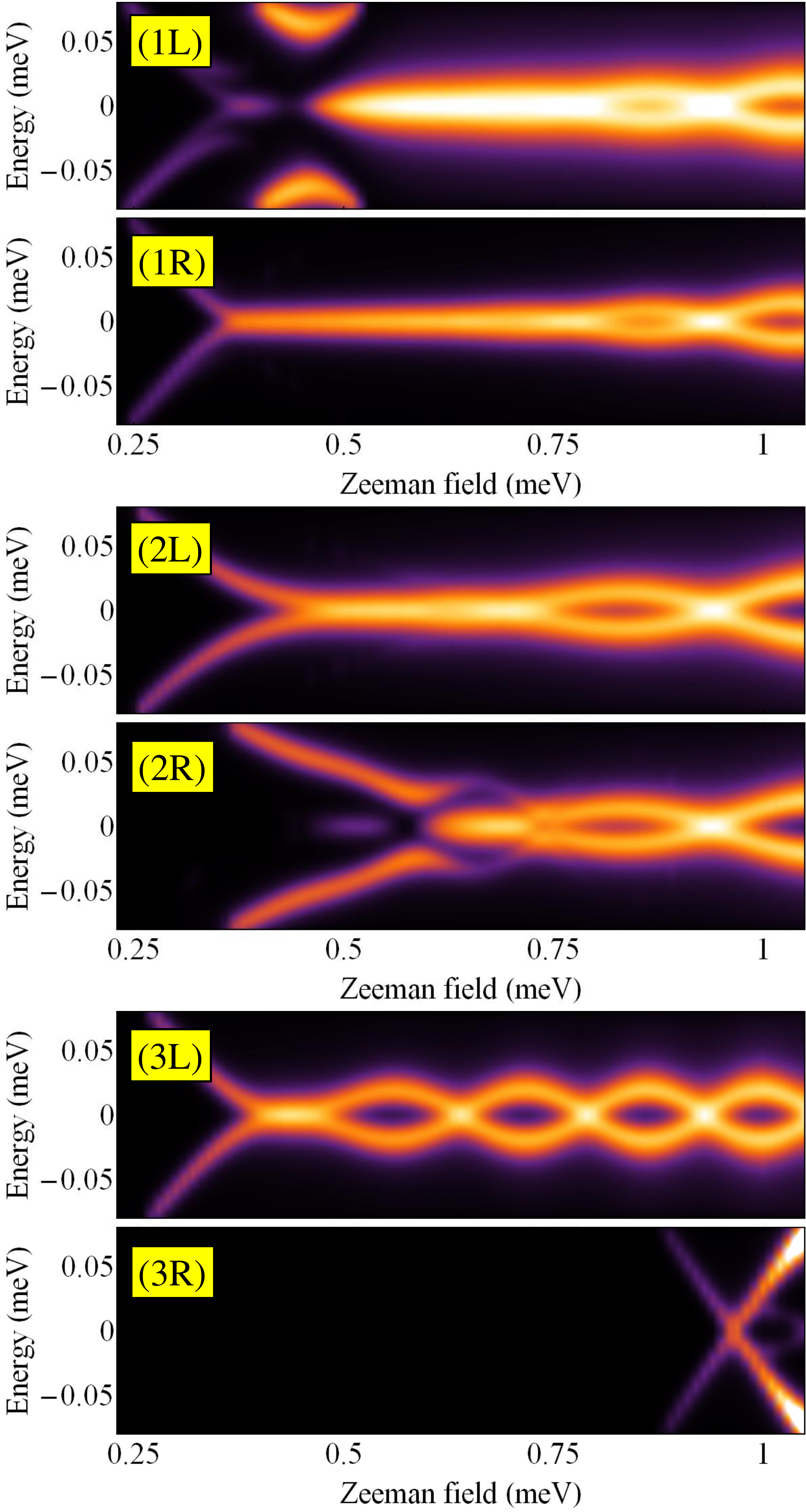}
\end{center}
\vspace{-2mm}
\caption{(Color online) Differential conductance as function of the applied Zeeman field and bias potential (energy) for the parameters corresponding to the panels marked by (1) and (3) in Fig. \ref{FIG2} and (2) in Fig. \ref{FIG3}.  The traces labeled by `L' and `R' correspond to tunneling from the left and right leads, respectively.  The traces shown in the top panels are correlated for Zeeman fields larger than about $0.5~$meV, those in the middle panels show correlations above $0.7~$meV, while the traces in the lower panels are uncorrelated.}
\label{FIG4}
\vspace{-3mm}
\end{figure}

In Fig.~ \ref{FIG4}, the top two panels correspond to a chemical potential $\mu = -1.4\Delta$ (see Fig. \ref{FIG2}), which implies a maximum amplitude of the effective potential $V_{\rm eff}(x)-\mu$ (over the proximitized segment of the wire) of about $2.6\Delta=0.65~$meV. Hence, for $\Gamma< \Gamma^*\approx 0.65~$meV the system is ``effectively homogeneous'' and one expects correlated features. Indeed, from Fig.~\ref{FIG4} one can infer the existence of correlated features (i.e., a robust ZBCP and a couple of splitting oscillations) for Zeeman field above roughly $0.5~$meV. Note that the low-energy mode that has a minimum gap for $\Gamma\approx 0.45~$meV  (see Fig. \ref{FIG2}) is only visible from the left end.

The middle two panels in Fig.~\ref{FIG4} correspond to a chemical potential $\mu = -1.2\Delta$ (see Fig. \ref{FIG3}), hence $\Gamma^*\approx 0.7~$meV. For $\Gamma>\Gamma^*$ one can clearly see correlated features, while at lower values of the Zeeman splitting the left and the right spectra are uncorrelated. Finally, for the lower two panels we have $\mu = -4.3\Delta$ (see Fig. \ref{FIG2}), which corresponds to $\Gamma^*\approx 1.1~$meV. No correlation can be observed in the spectra. The  MBSs responsible for the oscillatory  low-energy mode are localized on the left side of the wire (one near the left end of the wire, which produces the trace marked `3L' in Fig. \ref{FIG4}, and the other further inside). The Andreev bound state that crosses zero energy at $\gamma\approx 0.9~$meV (see Fig. \ref{FIG2}) is localized inside the smaller potential minimum near the right end of the wire (see Fig. \ref{FIG1}) and is only visible from the right end (see the trace marked `3R' in Fig. \ref{FIG4}.

We conclude that homogeneous (or effectively homogeneous) systems with a topological MZM localized at each end of the wire will be characterized by correlated differential conduction spectra, including energy splittings, critical fields associated with the emergence of ZBCPs, and even finite energy features (in highly homogeneous systems). Such features are completely uncorrelated, however, if the ZBCPs are due to ps-ABSs. These conclusions remain valid even if the ps-ABSs arise in a lead-quantum dot-nanowire set up \cite{Hansen,Prada,Liu-Sau}, in clean ballistic proximitized nanowires devoid of long length scale potential variations.  Note that having different left and right tunneling barriers may result in different visibilities for various correlated  features. The correlation is given be the corresponding energy scales and not by the values of the differential conductance at the two ends. Of course, adjusting the potential barriers can help obtain similar visibilities for the left and right spectra as well.

\section{Summary and Conclusion}
\label{Conclusion}

We have shown that the concept of {\em partially separated Andreev bound states} (ps-ABSs) is  useful in acquiring physical insight into the low-energy physics of semiconductor-superconductor hybrid structures and interpreting the experimental results. This concept  interpolates continuously between the ``garden variety'' ABS, which consists of two MBSs that overlap strongly in space, and the topological MZMs, which are separated by a distance given by the length of the wire ($L$). Strong spatial overlap between the constituent MBSs $\chi_A$ and $\chi_B$ implies that the energy of the corresponding standard ABS, as well as the corresponding signature in a tunneling conductance experiment, are extremely sensitive to the control parameters, e.g., Zeeman field, chemical potential, induced gap, and the height of the tunnel barrier.  By contrast, topological MZMs separated by a distance $\sim L >> \xi$ ($L$ being the length of the wire and $\xi$ the characteristic Majorana decay length) are characterized by an (exponentially) small  energy splitting and a quantized ZBCP that are extremely robust to perturbations. The ps-ABS interpolates smoothly between these two cases, but occurs on the topologically trivial side of the TQPT.  The ps-ABS corresponds to a pair of weakly coupled MBSs (with  a spatial separation between $\chi_A$ and $\chi_B$ of the order of $\xi$ or larger), which  results in a robust low-energy mode that ``sticks'' to near-zero-energy in an extended range of Zeeman field (and other control parameters) and generates quantized ZBCPs similar to the ``genuine'' Majorana zero modes (MZMs) localized at the ends of the wire.

The standard ``garden variety'' zero-energy Andreev resonance can be easily distinguished from MZMs by examining its stability to controllable external parameters (e.g., Zeeman field). By contrast, the possibility of the occurrence of robust near-zero-energy ps-ABS in an extended range of Zeeman field (Fig.~\ref{XFig2} lower panel and Fig.~\ref{fig:FFig4} lower panel) in the topologically trivial phase  (i.e., $\Gamma < \Gamma_c$ where $\Gamma_c$ corresponds to the bulk gap minimum), should give one a serious pause when interpreting the zero bias conductance peaks ubiquitously present in the existing tunneling conductance data \cite{Mourik_2012,Deng_2012,Das_2012,Churchill_2013,Finck_2013,Marcus,Hansen,HZhang,Frolov,Nichele,Quantized-ZBCP} in terms of topological MZMs. We illustrate the possible ps-ABS scenario by considering two relevant experimental SM-SC heterostructures, a lead-quantum dot-nanowire-superconductor junction \cite{Hansen,Prada,Liu-Sau} and a SM-SC set-up with long length scale potential variations in the proximitized nanowire.\cite{Stanescu-Tewari-2} In these calculations, the trivial low-energy ps-ABSs not only stick to near-zero energy over a wide range of Zeeman fields in the topologically trivial phase, but they are also equally robust against the variations of other controllable parameters, such as barrier height, chemical potential, and the induced superconducting gap, much as the topological MZMs in the topologically non-trivial phase. In addition, these ps-ABSs generate quantized zero-bias conductance peaks (ZBCPs) that are indistinguishable from the corresponding signatures of genuine MZMs.
In a recent theoretical work,\cite{Liu-Sau} it was claimed that the ZBCP due to ABS in the topologically trivial phase of the lead-quantum dot-nanowire-superconductor heterostructure is not robust against variation in the potential in the quantum dot and can have a quantized value only accidentally. These properties could be used to discriminate between low energy ABS and MZMs in this type of hybrid systems. However, we show that, in general, ps-ABSs with  constituent MBSs  partially separated are possible even in quantum dot-proximitized wire systems. This type of trivial low-energy  modes cannot be distinguished from true  MZMs using \textit{any} type of local measurement at the end of the wire. In particular, a tunneling measurement at the end of the wire results in a quantized zero bias conductance plateau insensitive to variations in the external parameters.\cite{Quantized-ZBCP}

In light of these results we conclude
that the observation of robust ZBCPs, even a ZBCP that has the ``expected'' quantized value of $2e^2/h$ at low temperature,  or the observation of  a zero-bias ``teleportation'' peak consistent with an
exponential decay of the energy splitting \textit{do not} represent unique signatures of topologically protected MZMs, because similar signatures can also appear in the effectively topologically trivial phase in the presence of ps-ABSs.
The ps-ABSs, on the other hand, cannot be used to demonstrate non-Abelian statistics or to perform topological quantum operations, which require  well separated MZMs localized at the ends of the wire.
We have shown that, in the absence of an interferometric measurement, the simplest and most straightforward way of discriminating between ps-ABSs and topological MZMs (where by ``topological MZMs'' we mean Majorana zero modes that are localized at the ends of the wire and can be used in TQC) is to perform separate tunneling measurements at the two ends of the wire and look for correlations. The absence of correlations is indicative of  ps-ABSs, while the presence of correlations is consistent with topological MZMs. We note that the only possibility of getting a ``false signal'' is  when identical quantum dots are present at both ends of the wire, or when the long length scale potential variations are reflection symmetric about a plane perpendicular to the wire. However, in this case the correlations are not robust against local variations of the control parameters (e.g., the gate voltage at one end of the wire).
\begin{acknowledgments}
CM and ST acknowledge support from ARO Grant No: (W911NF-16-1-0182). TDS was supported by NSF, DMR-1414683.
\end{acknowledgments}


\end{document}